\documentclass[epj]{svjour}
\usepackage{graphics}
\begin{document}
\title{A Theory of Pattern Recognition for the Discrimination between Muon and Electron in the Super-Kamiokande}
\titlerunning{A Theory of Discrimination in SK}

\author{V.I.~Galkin\inst{1} \and A.M.~Anokhina\inst{1}
\and E.~Konishi\inst{2}\and A.~Misaki\inst{3}
}                     
\offprints{}          
\institute{Department of Physics, Moscow State  University,Moscow, 119992, Russia
 \and Graduated School of Science and Technology,
Hirosaki University, 036-8561, Hirosaki, Japan
 \and Advanced Research Institute for Science and Engineering, Waseda University, 169-0092, Tokyo, Japan\\
 e-mail:misaki@kurenai.waseda.jp
}
\date{Received: date / Revised version: date}
%

%
\abstract{
The standard Super-Kamiokande analysis uses an estimator for
particle identification by which it discriminates 
$e$ ($\nu_e$) from $\mu$($\nu_{\mu}$).  Use of this estimator 
has led to the claim of a
significant deficiency of $\mu$ ($\nu_{\mu}$), suggesting the
existence of neutrino oscillations. We investigate three areas of
concern for the Super-Kamiokande estimator: the separation of the
spatial part from the angular part in the probability functions,
 the neglect of fluctuations in the Cherenkov light 
in different physical processes due to the charged 
particles concerned, and the point-like approximation for the
 emission of Cherenkov light.  We show that the first two 
factors are important for the consideration of stochastic
processes in the generation of the Cherenkov light, and that the
point-like assumption oversimplifies the estimation of the
 Cherenkov light quantities.  We develop a new discrimination
 procedure for separating electron neutrinos from muon neutrinos,
 based on detailed simulations carried out with GEANT~3.21 and
 with newly derived mean angular distribution functions for the
 charged particles concerned (muons and electrons/positrons),
 as well as the corresponding functions for the relative
 fluctuations.  These angular distribution functions are
 constructed introducing a ``moving point'' approximation.
 The application of our procedure between the discrimination
 between electron and muon to the analysis of the experimental
 data in SK will be made in a subsequent paper.
} 
\maketitle
\section{Introduction}
\label{intro}
The possible existence of neutrino oscillations is one of the most 
important issues in particle astrophysics as well as elementary 
particle physics at the present time. 
Among the positive and negative results reported for neutrino 
oscillation, experimental results for atmospheric neutrino by 
Super Kamiokande (hereafter, we abbreviate simply SK) has special position 
in the experiments concerned, because it is said that they have 
given the decisive and clear evidence 
for the existence of neutrino oscillation.
The reasons as follows: \newline
\indent (1) They carried out the calibration experiments 
for the discrimination between muon and electron by electron 
accelerator beam whose energies are well known and established 
the clear discrimination between muon and electron 
for SK energy region concerned \cite{Kasuga1}. \par
  
(2) Based on the well established discrimination procedure between 
muon and electron, they have analyzed {\it Fully Contained Events} 
and {\it Partially Contained Events}, whose energies covered from 
several hundreds MeV to several GeV.  As the results of them, 
they have found significantly different zenith angle distribution 
between for muon and electron, namely, muon deficit and 
attributed such discrepancy to  the neutrino oscillation between 
muon and tau.
As the most new one, they give   ${\rm sin}^2{2\theta} > 0.92$ and 
$1.5\times 10^{-3}{\rm eV}^2 < \Delta m^2 < 3.4\times 10^{-3}{\rm eV}^2$ 
at 90\% confidence level \cite{Fukuda1}.
 \par
 (3) Also, they have analyzed {\it Upward Through Going Particle Events} 
 and {\it Stopping Particle Events}. Most physical events under such 
 category could be regarded as exclusively  the muon (neutrino)
induced events, not electron (neutrino) induced events, because the 
effective volume for muon is much larger than that for electron due 
to longer range of muon irrespective the discrimination procedure 
between muon and electron which is indispensable for the analysis for 
{\it Fully Contained Events} and {\it Partially Contained Events}.  
Also, in this case, they have given the same parameters for neutrino 
oscillation which are obtained in the analysis of 
{\it Fully Contained Events} 
and {\it Partially Contained Events} \cite{Fukuda1}. 

Through three different kinds of the experiment performed by SK, 
all of which are constructed upon the well established procedure, 
it is said that SK has given clear and definite evidence for existence
 for the neutrino oscillation.
 \par
  
 The analysis of {\it Fully Contained Events} and {\it Partially Contained 
 Events} is closely and inevitably related to the discrimination 
 procedure between electron and muon. 
Because the frequency of muon  events with some energy 
occurred inside the detector is nearly the 
 same as that of electron events unless ossillation exists and,
 therefore, the precise discrimination procedure between
 electron and muon is absolutely necessary.

   Considering the great impact of SK experiment over other experiments 
concerned and theoretical physics, we feel we should examine the 
validities of the experimental results performed by SK, because 
nobody has examined them in the most comprehensive way, solely 
due to character of huge experiment, although the partial aspect 
of SK had been examined in fragmental way \cite{Olga}.

   However, Mitsui et al have examined the validity of the 
discrimination procedure by SK and have pointed out the necessity 
of fluctuation effect into the discrimination procedure 
between muon and electron  by SK \cite{Mitsui}.

   We have examined validities of all the SK experiment, adopting 
quite different approach from the SK procedure.


 In order to interpret the detected events
(i.e. to define the CAUSES for the CONSEQUENCES) one
should first go from CAUSES to CONSEQUENCES, i.e.
solve a DIRECT problem, and then back from
CONSEQUENCES to CAUSES, thus solving an INVERSE
problem. SK solve the DIRECT problem in a rather
simplified way and mostly concentrate on the INVERSE
problem. We go both ways with reasonable care.

  For the purpose, we have performed computer numerical experiments 
for examination of the validities of SK experimental results, 
constructing the virtual SK detector in the computer, 
the scale of which is same as the real SK detector.\newline
 
Concretely speaking, firstly, we have constructed the virtual small 
scale of SK detector, the scale of which is the same as the real 
SK detector (small scale) for being exposed to the accelerator beam, 
in the computer,for testing the validity of the discrimination between
 muon and electrons adopted by SK.
\footnote{The detailed study for the discrimination between 
electron and muon by using accelerator was carried out in Kasuga 
\cite{Kasuga2},\cite{Kasuga3} and Sakai\cite{Sakai}.} 
SK utlize their results by simple exterpolation.\par
Based on the imformation obtained from the virtual small scale of 
the detector,we have constructed the full size scale of the virtual 
SK detector, the size of which is the same as that of the real 
SK detector in the computer and have really checked whether the 
clear discrimination which SK assert is possible or not. 
In this case, we take account of the difference in the size between the pilot 
detector for the accelerator beam and the real (full size) detector.   
The methodology adopted by us is described in the present paper. 
In a pair of subsequent papers, we really compare our 
results under our methodology (present paper) with SK results. 
Further, the analysis of both {\it Fully Contained Events} 
and {\it Partially Contained Events} and the analysis of both 
{\it Upward Through Going Muon Events} and {\it Stopping Muon Events} 
in the numerical computer experiments will appear in the subsequent papers. 


An analysis of {\it Fully Contained Events} and 
{\it Partially Contained Events} is given in a subsequent paper.
An analysis of {\it Upward Through Going Muon Events} 
and {\it Stopping Muon Event} will be presented elsewhere. 

  From the methodological point of view for the analysis of the SK data, 
there should be two essential differences between the SK procedure and 
our procedure in the general analysis of physical events.

The first: as we already mentioned above, SK
oversimplifies the solution of the DIRECT problem,
namely, their mean models of both electron and muon
events are too far from reality to be used in the 
INVERSE problem solution, while our mean models are
reasonably accurate.
The second: SK neglects fluctuation effects in physical processes 
in which the Cherenkov light is produced, while we consider them 
as correctly as possible, which we discuss in the section 2 and 
the subsequent sections.

  In this paper, we limit our discussion to the discrimination 
procedure between muon and electron in the SK experiment 
(the item(1) in the Introduction)\\

 The contens of the paper are organized as follows: In the section 2, 
we examine the SK standard discrimination procedure in detail. 
We examine the point-like approximation in the electron shower
due to electron neutrino adopted by SK and conclude that such 
approximation leads to serious error in the energy determination 
of the electron neutrino. Next, we examine the probability 
functions for the pattern part and the angular part in both 
muon and electron and conclude that such separation of the pattern 
part from the angular part is not adequate and further, there are, 
a priori, no reasons why the probability functions  for muon obey 
the same type of that for the electron, even if such separation is valid.

There are many factors which produce various errors in the SK 
discrimination procedure. The simple agebraic sum of such errors 
not always denote right resultant errors.  In our discrimination 
procedure, we examine the resultant errors from the SK standard 
discrimination procedure, for example, the error in the vertex 
points of the neutrino interactions and their directions and more 
precise values.

 In the section 3, we develope more suitable discrimination 
procedure instead of the SK standard procedure and,
 for the purpose, costruct the mean angular distribution 
functions for Cherenkov light due to muon and electron
 and the corresponding relative fluctuation functions
 which make it possible to estimate the degree of the 
separation between muon and electron, errors of the vertex points 
for muon-like event and electron like events, the error of the 
directions due to these events.
In a subsequent paper, we develop the general procedure for 
estimating the concrete errors for the neutrino events concerned 
and give finally various errors for the neutrino events 
quantitatively by comparing our resluts with SK results.


\section{Examination of the SK discrimination procedure between 
muon and electron}
\label{sec:2}
   The approach for the discrimination between muon and
 electron adopted by SK is as follows.

\indent(1)SK calculate Cherenkov images for electron
and muon events by statistical and deterministic
numerical methods but while constructing the pattern
recognition procedure all fluctuation data and some
important features of the mean models are completely
ignored which are discussed below.
Such manipulation leads to serious error in both pattern recognition 
between electron and muon and their energy estimation.\\

\indent (2) Based on the Cherenkov light of the particle concerned 
mentioned above, SK constructed the probability function, 
ESTIMATOR for the particle identification, which is composed of 
pattern part and the angular part in the separated form.

The probability function for pattern,
$P_{pattern}(e(\mu))$,
which will appear in the section 2.2, denotes the pattern 
of the events concerned, namely, spatial and angular image of the 
events concerned. However, the variety of the images of the events 
concerned in SK discrimination procedure exclusively come from the 
fluctuations in the photoelectrons in the PMT, but not from the 
physical processes of the events concerned which produce the 
Cherenkov light, while in our procedure the image of the events 
concerned are essentially governed by the physical proccesses 
concerned which are completely neglected in the SK procedure.\\

The probability function for the angular part is introduced into 
the total probability (estimator for particle identification), 
being separated from that for pattern part.  This is  
unnatural, however, because the 
pattern is of the concept on the space-angular structure.
 And it is natural that 
the concept of the angular should be included in the concept 
of the pattern.
In our procedure, we construct the concept of the pattern as the 
angular-space image in which the spatial one is interrelated with the 
angular one (See, section 3.3).   

Further, SK adopt the same type of the probability functions for 
muon and electron. However, there are no reasons why the probability 
function for muon obeys that for electron, because the generation 
mechanism of the total Cherenkov light is quite different in muon 
from in electron.\\ 
In present paper, we would emphasize the neglect of the fluctuation 
effects and the oversimplification in the problems adopted by SK 
lead to serious errors on the discrimination between muon and electron, 
which are  shown in the present section and subsequent sections.


\subsection{The Cherenkov light calculation for electron and muon.}
\label{sec:2.1}

\subsubsection{Examination on the Treatment of Cherenkov light for 
electron by SK}
\label{sec:2.1.1}
  The SK analysis approximates the sources of the Cherenkov light 
due to electron showers as being point. 
In the calculation of the source for the Cherenkov light due to 
shower particles, the SK analyze electron events in average values, 
neglecting fluctuation and further assigns them to be point-like source.
Such a treatment is not valid for the analysis of the real physical 
events concerned in the SK detector, because the sources for the 
Cherenkov light due to shower particles definitely have
 geometrical extents. 
The geometrical extent of electron cascade shower as a light
 source is not negligible compared with the scale of the SK
 detector.

Thus, the introduction of the point-like approximation into the 
construction of the ESTIMATOR for particle identification
 generally invites not negligible error into the analysis of
 experimental events.
In what follows, we examine the limitations of the  point-like 
approximation by SK.
 Here, we restrict our examination to the direct Cherenkov light, 
because the scattered Cherenkov light is not essential for the 
examination of the logical structure of the SK discrimination 
between electron and muon and we are exclusively interested in 
the logical structure of the SK procedure.\par

In order to construct the ESTIMATOR for electron identification, 
the SK calculate the mean number of photoelectrons which are produced by 
the Cherenkov light due to an electron-primary cascade shower 
(electron neutrino) using the Monte Carlo method. Thus, SK obtains, 
$N_{i,exp}(direct)$, the expected number of photoelectron from the 
direct Cherenkov light initiated by an electron in the $i$~th PMT at 
a distance and an angle, in the following : 
\cite{Kasuga3}, \cite{Sakai}, \cite{Kibayashi}

\begin{eqnarray}
\lefteqn{N_{i,exp}(direct)= } \nonumber \\
&&=
\alpha_e \cdot N_MC(\theta_i,p_e)
\cdot \left( \frac{16.9}{l_i} \right)^{\gamma} \cdot \mbox{exp}
 \left( -\frac{l_i}{L} \right)\cdot f(\Theta),
\label{eqn:1}
\end{eqnarray}
\noindent
where $N_{MC}(\theta_i,p_e) $  denotes the mean number of the 
photoelectrons received in a circular area of 50 cm diameter 
located on a sphere of 16.9 meter in radius by the Full Monte 
Carlo simulation, $\alpha_e$  is the normalization factor, 
$l_i$  is the distance from the particle position to the $i$~th PMT, 
$\theta_i$ is the angle of the $i$ th PMT from the particle direction, 
and  $f(\Theta)$ is the effective photo-sensitive area of the PMT.
 Here, $L$ is the attenuation length of the Cherenkov light in the 
 SK detector, which is taken 100 meter. 
The third term in the right-hand side of Eq.(1) shows the
 attenuation of light 
and its power index is estimated from Monte Carl simulation. 
The value of $\gamma$ is 2.0 in Takita (p83, in \cite{Takita}) and 
 Kasuga(p32 in \cite{Kasuga2}) and Sakai(p41 in \cite{Sakai}), 
and 1.5 in Kasuga(p71 in \cite{Kasuga3}). 
A numerical value of 2.0 is used in older analysis while 1.6 is 
adopted in more recent work.
The expression of Eq.(1) is an oversimplification 
on the estimation of photoelectrons 
by the Cherenkov light when we consider the real behaviour
of the electron cascade shower. 
In order to clarify the oversimplification of the SK procedure,
 let us compare it with our procedure in the level of 
 Cherenkov light, but not in the photoelectron level,
 because it is enough for the purpose concerned.
Here, we coulud define the following expression which
correspond to Eq.(1),
\begin{eqnarray}
\lefteqn{N_{exp,cher,app}(direct)= } \nonumber \\
&&=
\alpha'_e \cdot N_{MC,cher}(\theta, p_e)
\times \left( \frac{16.9}{l} \right)^{\gamma} \times \mbox{exp}
 \left( -\frac{l}{L} \right),
\label{eqn:2}
\end{eqnarray}

\noindent where $N_{MC,cher}(\theta,p_e)$ is the mean
Cherenkov photon density which correspond to
$N_{MC}(\theta_i,p_e) $ in Eq.(1), $\alpha'_e$ is another
normalization factor and other parameters are same as in Eq.(1). 
Namely, Eq.(2) gives
 the Cherenkov photon density for the electron cascade 
shower under the point-like 
approximation which is utlized by SK.
The dependence of the Cherenkov photon density 
$N_{exp,cher,app}(direct)$ on the distance $l$
is essentially determined by  $(16.9/l)^{\gamma}$
 rather than exp(-$l/L$),
because the attenuation of the Che-\\renkov light  
is small (100 meter of attenuation length in SK). 
Thus, the Cherenkov light depends strongly on the inverse power of the distance in the SK procedure. 
\begin{figure}
\begin{center}
\hspace*{-0.5cm}
\rotatebox{90}{%
\resizebox{0.4\textwidth}{!}{
  \includegraphics{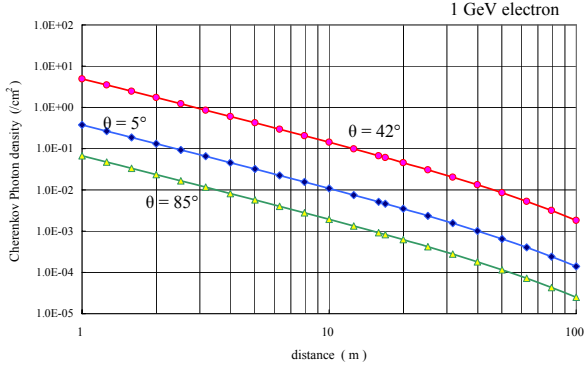}
}}
\vspace{-1.5cm}
\caption{\label{fig:1} The dependence of the Cherenkov photon 
density due to the electron shower on the distance for differnt
 angle from the direction of an  incident 1 GeV electron with
 the point-like approximation, normalized with the values at
16.9m in Figure~2. The angles considerd are 5, 42 and 85 degree. 
Sampling numbers of Monte Carlo simulations 
are 10000 per each angle. See the text for detail.} 
\end{center}
\end{figure}

\begin{figure}
\begin{center}
\hspace*{-0.5cm}
\rotatebox{90}{%
\resizebox{0.4\textwidth}{!}{
  \includegraphics{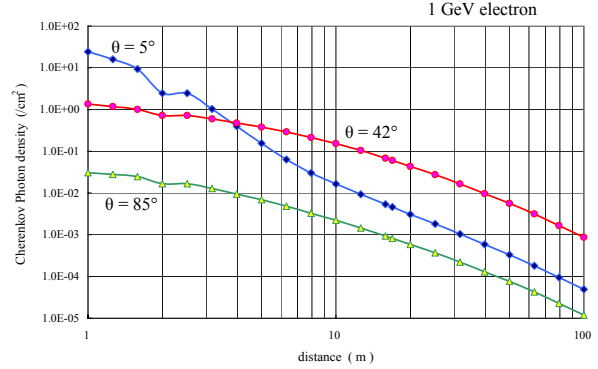}
}}
\vspace{-1.5cm}
\caption{\label{fig:2}  The dependence of the Cherenkov photon 
density due to the electron shower on the depth for differnt 
angle from the direction of an  incident 1 GeV electron
 without the point-like approximation. The angles 
considerd are 5, 42 and 85 degree. Sampling numbers of Monte Carlo simulations 
are 10000 per each angle. See the text for detail.}
 
\end{center}
\end{figure}

In Figure~1, the dependence of the Cherenkov photon 
density due to the electron cascade shower on the distance ($l$)
 are given
for differnt angle from the direction of an incident 1 GeV
 electron with the point-like approximation using Eq.(2).
The energy value of 1 GeV is a typical energy in the SK detector.

In order to examine the validity of the point-like 
approximation by SK, we have carried out the calculation of
 $N_{exp,cher,FMC}(direct)$, 
the Cherenkov photon density without the point-like 
approximation in Eq.(3) for the comparison with Eq.(2).
Instead of the point-like approximation, we simulate the
 electron cascade shower exactly in the stochastic manner,
 which we call [without point-like approximation].
 Then, the Cherenkov photon density generated by the shower
 elctron is calculated by help of GEANT3.21.

  Namely, Eq.(3) denote the straightforward expression 
for the Cherenkov light density which is directly derived
from the electron cascade shower without the approximation of
$(16.9/l)^{\gamma}$ and exp(-$l/L$).
  Thus, the definition of [ without point-like approximation] 
denotes that we simulate electron cascade shower and accompanied
Cherenkov light density in three-dimensional way at any distance
and angle as exactly as possible. Namely, we calculate the whole
structure of the Cherenkov photon density, taking into account
the attenuation of light, as well as that of the electron cascade
shower. Then, we obtain
\begin{equation}
N_{exp,cher,FMC}(direct)=N_{MC,cher,FMC}(\theta,p_e,l),
\label{eq:3}
\end{equation}
\noindent where $N_{MC,cher,FMC}(\theta,p_e,l)$ 
denotes the Cherenkov 
photon density at the distance $l$ without introducing the
 point-like approximation, 
which is obtained by the Monte Carlo method, taking into account 
the attenuation length of the Cherenkov light correctly.

In Figure~2, the dependence of the Cherenkov photon density
 on the distance without point-like approximation is given for
 the different angles from the direction of the incident 
elctron using Eq.(3).
In [without point-like approximation], the dependence on the 
distance and angle and the attenuation 
are automatically included into functions themselves.  

From the Figure~1, it is clear that the dependences of the 
Cherenkov photon density on the distance in the SK are same irrespective
 of the angles, which simply reflects the separation of the
 distance part from the angular part in Eq.(2).  
Comparing the Figure~2 with Figure~1, 
it is easily understood
 that in the region beyond about 10 meter, the point like 
approximation 
is well approximation as far as the tendency in concerned,
except the absolute values, while  in the region inside about 
10 meter, this approximation does not hold anymore, due
to the actual longitudinal and lateral structure of 
the electron cascade showers.
 In particular, for direction 
within 5 degree, the effect of the lateral spread of 
the electron cascade shower as well as that of the 
definite longitudinal length becomes effective and the 
actual dependence on the distance is deviated largely 
from that in the point-like approximation.   

  In Figure~3 and Figure~4, we give the angular dependence 
of the Cherenkov photon density for different distances 
[with point like approximation](Eq.(2)) 
and [ without point like approximation ](Eq.(3)), respectively. 
 In Figure 3, the shapes of their angular dependence is exactly
same irrespective of their distance, because their angular
part is separated from their distance part, as shown in Eq.(2).
 Such separation of the angular part from the distance part does
not reflect real situation of the Cherenkov light density for 
  electron, as shown in Figure~4.

In Figure~4, we give the corresponding ones to Figure~3 
in the case of without point-like approximation.
 It should be noticed that for the region of smaller than 5~m 
the tendency of the angular dependence on angle is largely deviated from that for the region of larger distances.
  This reflects the fact that the Cherenkov photon density is
 largely influenced by the longitudinal and lateral structure
 of the real electron cascade shower.  
From figure~1 to 4 normalization are made at 16.9 meter.
\begin{figure}
\begin{center}
\hspace*{-0.5cm}
\rotatebox{90}{%
\resizebox{0.4\textwidth}{!}{
  \includegraphics{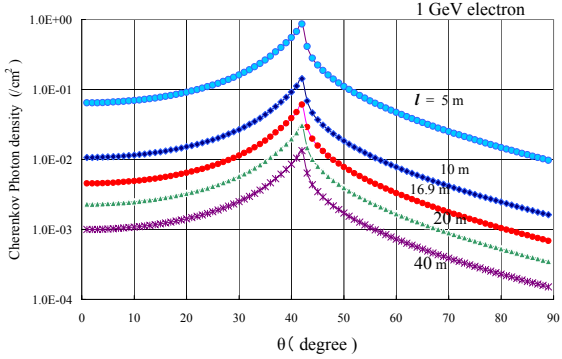}
}}
\vspace{-1.5cm}
\caption{\label{fig:3} Cherenkov photon density as a function
 of angle for different distances with the point-like 
approximation normalized with the values at
16.9~m in Figure~4 for 1 GeV primary electron.}
 \end{center}
\end{figure}
\begin{figure}
\begin{center}
\hspace*{-0.5cm}
\rotatebox{90}{%
\resizebox{0.4\textwidth}{!}{
  \includegraphics{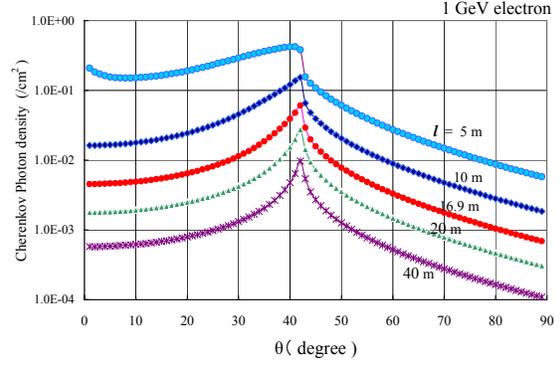}
}}
\vspace{-1.5cm}
\caption{\label{fig:4}  Cherenkov photon density as a function
 of angle without the point-like approximation for 1 GeV primary
 electron. Other parametersare the same as in Figure~3.}
 
\end{center}
\end{figure}

\begin{table}
\begin{center}
\vspace{5mm}
\caption{\label{table_1a}
The ratio of the Cherenkov photon density from electron showers
 with the under point-like approximaton (Eq.(2),$\gamma=1.5$) 
to corresponding ones without the point-like approximation
 (Eq.(3)). The primary enegy of electron is 1~GeV.}
\begin{tabular}{cccccc}
\hline
	angle &&&distasnce&&(m)\\
	(degree)&5.00&10.00&16.90&20.00&40.00 \\
\hline
5&0.175&0.553&1.000&0.948&1.452\\ 
10&0.191&0.548&1.000&0.949&1.457\\ 
20&0.200&0.525&1.000&0.956&1.494\\ 
30&0.231&0.495&1.000&0.973&1.583\\ 
40&0.549&0.640&1.000&0.939&1.505\\ 
50&0.757&0.756&1.000&0.887&1.135\\ 
60&0.715&0.739&1.000&0.893&1.166\\ 
70&0.708&0.736&1.000&0.893&1.171\\ 
80&0.708&0.737&1.000&0.892&1.167\\ 
85&0.706&0.737&1.000&0.891&1.164\\ 
\hline
\end{tabular}
\end{center}
\end{table}

\begin{table}
\begin{center}
\vspace{5mm}
\caption{\label{table_1b}
The ratio of the Cherenkov photon density from electron showers
 with the  point-like approximaton (Eq.(2),$\gamma=2.0$) 
to corresponding ones  without the point-like approximation
(Eq.(3)).  The primary enegy of electron is 1~GeV.}
\begin{tabular}{cccccc}
\hline
	angle &&&distasnce&&(m)\\
	(degree)&5.00&10.00&16.90&20.00&40.00 \\
\hline
5&0.322&0.719&1.000&0.872&0.944\\ 
10&0.352&0.712&1.000&0.872&0.947\\ 
20&0.368&0.682&1.000&0.879&0.971\\ 
30&0.425&0.643&1.000&0.894&1.029\\ 
40&1.009&0.832&1.000&0.863&0.978\\ 
50&1.393&0.983&1.000&0.815&0.738\\ 
60&1.314&0.960&1.000&0.821&0.758\\ 
70&1.301&0.956&1.000&0.821&0.761\\ 
80&1.301&0.958&1.000&0.820&0.758\\ 
85&1.299&0.958&1.000&0.819&0.757\\
\hline
\end{tabular}
\end{center}
\end{table}

In Table~1, we give the ratio of the Cherenkov photon density 
with the point like approximation to that without the point-like
 approximation for the exponent of 1.5(Kasuga,p74\cite{Kasuga3}). 
In Table 2, 
 we give the same quantities for the exponent 2.0, the older value
 used in the SK analysis(Takita,p83\cite{Takita}, Sakai,p41\cite{Sakai}). 
The ratios are normalized to the values in Eq.(3) at the 16.9 meter in the both cases.

Comparing Table~1 and Table~2, it is clearly understood that the 
ratios depend strongly on the values of $\gamma$.
  If the point like 
approximation is valid, then, the ratios in Table 1 and Table 2 
should remain around 1.0 with satisfactory allowance for every 
distance and every angle. However, it is clear from the tables 
such situation is never realized.  SK guarantee the accuracy of
 the energy determination for the event is within $\pm$2.6 \%
(Ishitsuka $\pm$2 \% (p29 in \cite{Ishitsuka}),
Kameda $\pm$2.5 \% (p62 in \cite{Kameda}),
Okumura $\pm$2.5 \% (p45 in \cite{Okumura}),
Messier $\pm$2.5 \% (p92 in \cite{Messier})
and Kasuga $\pm$2.6 \% (p53 in \cite{Kasuga3})).
 This is quite far from the reality.   In conclusion, Eq.(1), 
 the expression for the photoelectron utilized by SK, does not
  consider  the longitudinal and lateral structure of the electron
 cascade and, therefore, an oversimplification which could not 
 estimate the photoelectrons by the Cherenkov light correctly. 


\subsubsection{Cherenkov light for muon}
\label{sec:2.1.2}
  In contrast to the case of electron, SK calculates the Che-\\
renkov light,
taking into account the definite extent of muon range, namely, without 
point-like approximation. However, still, SK neglect the fluctuation effect.
  Here, let us examine the validity of the expression adopted by the 
standard SK analysis for a muon event, in which the expected number
of photoelectrons in the $i$~th PMT produced by a muon is
 expressed as Takita(p84 in \cite{Takita}),\par 
Sakai(p41 in \cite{Sakai}), Kasuga(p74 in \cite{Kasuga3}),
 Kibayashi(p71 in \cite{Kibayashi}),
\begin{eqnarray}
\lefteqn{N_{i,exp}(direct)= } \nonumber \\
&=&
\left\{ \alpha_{\mu} \times \frac{1}{l_i(sin\theta_i+l_i 
\times (\frac{d\theta}{dx}))} \times sin^2 \theta_i + N_{i,knock}(\theta_i) 
\right\} \nonumber \\
\lefteqn{\times\mbox{exp} \left( -\frac{l_i}{L} \right)
\times f(\Theta),}
\label{eqn:4}
\end{eqnarray}

where  $\alpha_{\mu}$  is the normalization factor. The second term 
in the right-hand side originates from the ionization energy loss 
dE/dx in water. Taking into account the change in the photon density 
which is caused by the change in, corresponding to the energy loss, 
$dx sin \theta + l d \theta$, expresses the intensity variation 
of the Cherenkov photons.\par
  $N_{i,knock}(\theta_i) $  shows the number 
of expected photons from knock-on electrons as a function of, 
which is estimated by a Monte Carlo simulation.
  It is easily understood from Eq.(4) that the logic for producing
photoelectron in the muon  is definitely deterministic.  
In other words, the SK analysis neglects fluctuation 
effects on the generation of the Cherenkov light from muon completely.
 However, the muon losses its energy and changes the direction 
 in stochastic way, namely, knock-on, 
the decay-product electron and multiple scattering. 
Therefore, we could not neglect fluctuations for the muon event. 
As one example of muon behavior, we show the range fluctuations for 
muon in Figure~5. The neglect of fluctuations for muon causes 
uncertainty in the estimation of the muon energy.

\begin{figure}
\begin{center}
\resizebox{0.3\textwidth}{!}{%
  \includegraphics{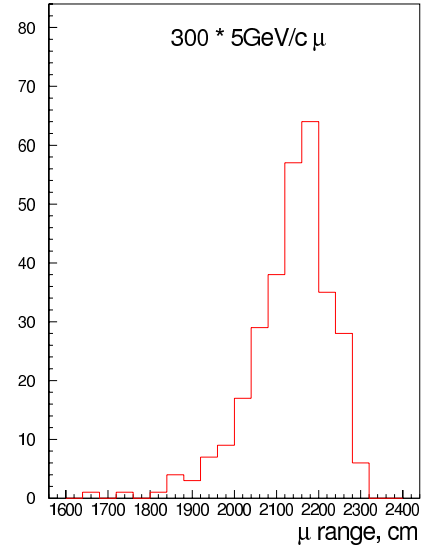}
}
\caption{\label{fig:5} Range fluctuation of 5 GeV muons. The 
number of simulation is 300.}
\end{center}
\end{figure}


\subsection{Pattern recognition procedure adopted by SK}
\label{sec:2.2}
  In previous subsection, we examine accuracies on the Che-\\renkov 
light quantities for electron and muon, which is related to the 
estimation of the energies of the particles concerned directly
for the Fully Contained Events.
In the present section, we should like to say these quantities are 
of major importance for the type definition and geometry recognition, 
both in SK and our case. 
Here, we examine pattern recognition procedure for the particle concernrd. 


\subsubsection{Principle of the ESTIMATOR of particle 
identification adpted by SK}
\label{sec:2.2.1}

  SK defines the probability, $Prob_i(N_{exp},N_{obs})$, suitable for 
  the particle identification. 
It is the probability to observe the number of photoelectron, $N_{obs}$, 
for an expected mean number of photoelectron, $N_{exp}$. The $N_{obs}$
actually deviates from the ideal Poisson distribution due to 
fluctuations in amplification processes of the PMTs. 
The probability function is simply given by:

\begin{equation}
Prob=\frac{1}{\sqrt{2\pi}\sigma}\mbox{exp} \left( -\frac{(N_{obs}-N_{exp})^2}
{2\sigma^2} \right),   
\label{eq:5}
\end{equation}

\noindent
where $Prob_i$  is the probability function for the $i$ th PMT.
Thus, SK define the likelihood functions for e-like event and mu-like 
event is the following:

\begin{equation}
\begin{array}{rcl}
L_e & = & \displaystyle \prod_{\theta_i<(1.5 \times \theta_C)}\!\!\! 
Prob_i(e), \\& & \\
L_\mu & = & \displaystyle \prod_{\theta_i<(1.5 \times \theta_C)}\!\!\! 
Prob_i(\mu), 
\label{eq:6}
\end{array}
\end{equation}
\noindent
where $Prob_i(e)$ is calculated assuming the event is due to electron. And
 $Prob_i(\mu)$ due to muon. In order to combine the information of the
 Cherenkov opening angle with the information of the ring pattern,
  $L_e$ and $L_{\mu}$  are transformed into  $\chi^2$ distribution 
  so that SK obtain the following functions for spatial part and 
  angular part.
\footnote{In our opinion, a priori, there are no reasons why does 
the probability function for muon obey
 the same type of the probability function for electron.}  
\begin{equation}
\begin{array}{rcl}
P_{pattern}(e) & = 
& \displaystyle \mbox{exp} \left\{ -\frac{1}{2}\left( \frac{\chi^2(e)-min
[\chi^2(e),\chi^2(\mu)]}{\sigma_{\chi^2}} \right)^2 \right\}. \\
& & \\
P_{pattern}(\mu) & = 
& \displaystyle \mbox{exp} \left\{ -\frac{1}{2}\left( \frac{\chi^2(\mu)-min
[\chi^2(e),\chi^2(\mu)]}{\sigma_{\chi^2}} \right)^2 \right\}. \\
\end{array}
\label{eq:7}
\end{equation}

\begin{equation}
\begin{array}{rcl}
P_{angle}(e) & = & constant \times \displaystyle \mbox{exp} \left
\{ -\frac{1}{2} 
\left( \frac{\theta_{exp}(e)-\theta_{obs}}{\Delta \theta} \right)^2 \right\} \\
& & \\
P_{angle}(\mu) & = & constant \times \displaystyle \mbox{exp} \left
\{ -\frac{1}{2} 
\left( \frac{\theta_{exp}(\mu)-\theta_{obs}}{\Delta \theta} \right)^2 \right\}
\end{array}
\label{eq:8}
\end{equation}
\\
Combining (6) with (7), SK obtains probability functions for electron 
and muon finally in the following:
\begin{equation}
\begin{array}{rcl}
P(e) & = & P_{pattern}(e) \times P_{angle}(e),  \\
& & \\
P(\mu) & = & P_{pattern}(\mu) \times P_{angle}(\mu).
\end{array}
\label{eq:9}
\end{equation} 
\\

 A ring is more mu-like than e-like, if  $P(\mu)>P(e)  $   and vice versa. 
  


\subsubsection{Examination of the components of the ESTIMATOR adopted by SK}
\label{sec:2.2.2}
   Let us examine the validity of Eq.(5) in the SK procedure. 
[1] They neglect fluctuation effects in physical processes for 
electron and muon.
 SK introduce Eq.(5) for the detection of the particle concerned. 
However, they consider only fluctuation effect coming 
from the amplification 
processes in the PMT, namely, electro-mechanical effect of the 
hardware detector, and never any fluctuation effect coming from the 
physical process for electron and muon, namely, fluctuation from 
shower particles for electron (neutrino) and range straggling of 
muon for muon (neutrino) which play essential role for formation 
of variety of  the pattern onto the detector for electron and muon. 
Namely, SK describe the behavior of electron and muon in the mean values. 
Consequently, varieties of the Cherenkov light pattern due to 
electron and muon are produced only through "Poison distribution 
in the amplification process in the PMT, the position of the neutrino 
interaction and the direction of the incident neutrino. 
However, essential character of the varieties of the Cherenkov 
light due to electron and muon are only produced through the 
fluctuation in the physical processes of the particle concerned.
If we consider fluctuation effect correctly, Eq.(1) defined by 
SK should be replaced by

\begin{eqnarray}
\lefteqn{P_e(N:E_0)= } \nonumber \\
&&=
P_e(N_{e,obs},N_{e,exp}(E_0)) \times P_e(N,N_{e,obs}),
\label{eqn:10}
\end{eqnarray}
\noindent
 where  $P_e(N:E_0)$ denotes the probability for an electron with energy
 $E_0$  to produce a cascade shower in which $N$ photoelectron are 
 produced by the Cherenkov light due to shower particles.    
$P_e(N_{e,obs},N_{e,exp}(E_0)) $ is the probability for an electron 
with primary energy $E_0$  and the mean number of the Cherenkov photon,  
$N_{e,exp}(E_0)$, to produce the observed number of the Cherenkov photons,
$N_{e,obs}$ and  $P_e(N,N_{e,obs})$  denotes the probability for $N_{e,obs}$,
number of the Cherenkov photons to produce finally the number of the 
photo-electron N.
$P_e(N_{e,obs},N_{e,exp}(E_0)) $   shows the fluctuation effect 
in a electron casade shower and  $P_e(N,N_{e,obs})$ correspond 
to Eq.(5) in SK procedure.
  For muons, the expression which correspond to Eq.(8) is given as follows:


\begin{eqnarray}
\lefteqn{P_{\mu}(N:E_0)= } \nonumber \\
&&=
P_{\mu}(N_{\mu +e,obs},N_{\mu +e,exp}(E_0)) \times P_{\mu}
(N,N_{\mu+e,obs}),
\label{eqn:11}
\end{eqnarray}
\noindent
where $P_{\mu}(N:E_0)$ is the probability for a muon of 
primary energy $E_0$   
to produce the number of the photoelectron N by the Cherenkov 
light due to the muon and its accompanying electron (knock on 
electron and others).
 $P_{\mu}(N_{\mu +e,obs},N_{\mu +e,exp}(E_0)) $ 
shows the fluctuation effect in the physical processes of muon, 
namely, the range struggling of muon.
Corresponding to Eq.(5), $P_e(N,N_{e,obs})$ and
$P_{\mu}(N,N_{\mu+e,obs})$  
may be approximated by Gausian distribution with different N. 
However, $P_e(N_{e,obs},N_{e,exp}(E_0)) $  
should be different from  
$P_{\mu}(N_{\mu +e,obs},N_{\mu +e,exp}(E_0)) $, 
because the physical process for producing the Cherenkov light 
for electron is clearly different from that for muon.  
Consequently,  there are no reasons, a priori, why\\ 
$P_e(N_{e,obs}, N_{e,exp}(E_0)) $ and $P_{\mu}(N_{\mu +e,obs},
N_{\mu +e,exp}(E_0)) $\\
should obey the same Gaussian distribution.
In conclusion, Eq.(9) lacks in theoretical background as the 
probability function in the practical application to the 
analysis of experimental data.





\subsubsection{Resultant errors introduced by SK discrimination procedure}
\label{sec:2.2.3}

  In previous subsection, we examine errors introduced into the 
Cherenkov light quantities for electron and muon. The final purpose 
of the construction of the ESTIMATOR adopted by SK is [a] the 
discrimination between electron and muon, [b] decision of the vertex 
position in the neutrino event and decision of the direction of the 
individual neutrino event and [c] the decision of the momentum of 
the particle concerned. Therefore, one may be very interested in the 
resultant errors which SK procedure invite due to inadequency of the 
ESTIMATOR, the neglect of the fluctuation effect and assumption of 
the point-like approximation.

  The answer is as  follows: we could not estimate individual error 
in the SK procedure, for example, the error due to the the 
oversimplification in the treatment of the electron cascade shower. 
The individual error is intermingled with each other so that we get 
the resultant errors which are never the simple sum of the individual error.  
Namely, we could get the resultant errors only, for example, the 
error of the energy measurement in the neutrino events, the errors on 
the vertex in the neurino event, the error on the direction of the 
neutrino even and so on.

  In the next section, we develop our rigourous procedure which 
does not include inconsistency of the logic.


\section{Principle of our discrimination procedure between electron and muon:
Application of pattern 
recognition theory to e-mu discrimination.}




\subsection{Overcome to the defect of the SK procedure}

  In the previous section, we clarify that the estimation of the 
Chrenkov light quantities  due to the electron event and the muon 
event as well as the discrimination of their pattern are not reliable.
Because, both electron events and muon events are recognized without 
taking their fluctuation effect which may kill the real variety of 
their pattern. 
In the electron events, the point-like approximation 
 may lead serious misestimation of their  
as shown in Table~1 and Table~2.
In order to overcome the defects inherent in the SK 
discrimination procesure, we develop our discrimination procedure 
as exaxtly as possible, taking the stochastic characters in the 
physical processes concerned carefully into account.

 Here, we restrict the development of our theory at the level of 
the Cherenkov light, but not at the photoelectron level which are
 directly related to the actual experimental result as 
in the SK procedure. The reasons are as follows:
 The most important of our
present paper is to clarify the essential defect adopted by SK and 
propose the direction of the possible alternative procedure
 by which SK procedure could be replaced.

In this case, it is enough for us to examine physical quantities
in the level of the Cherenkov light leaving out the consideration 
of photon conversion into photoelectron. Such simplification 
reduces the overall uncertainties of parameter estimates and, 
thus, our results give the lower limits for 
the errors in the analysis of the neutrino events concerned. To illustrate 
the scale of uncertainty added by the light-to-electron conversion, in Figure~6 
we give the probability distributions of the number of the Cherenkov photon 
for given number of photoelectron. Of course, while applying our methods to 
the analysis of the real experiment one should take photoelectron 
fluctuations into account.


\begin{figure}

\resizebox{0.5\textwidth}{!}{%
  \includegraphics{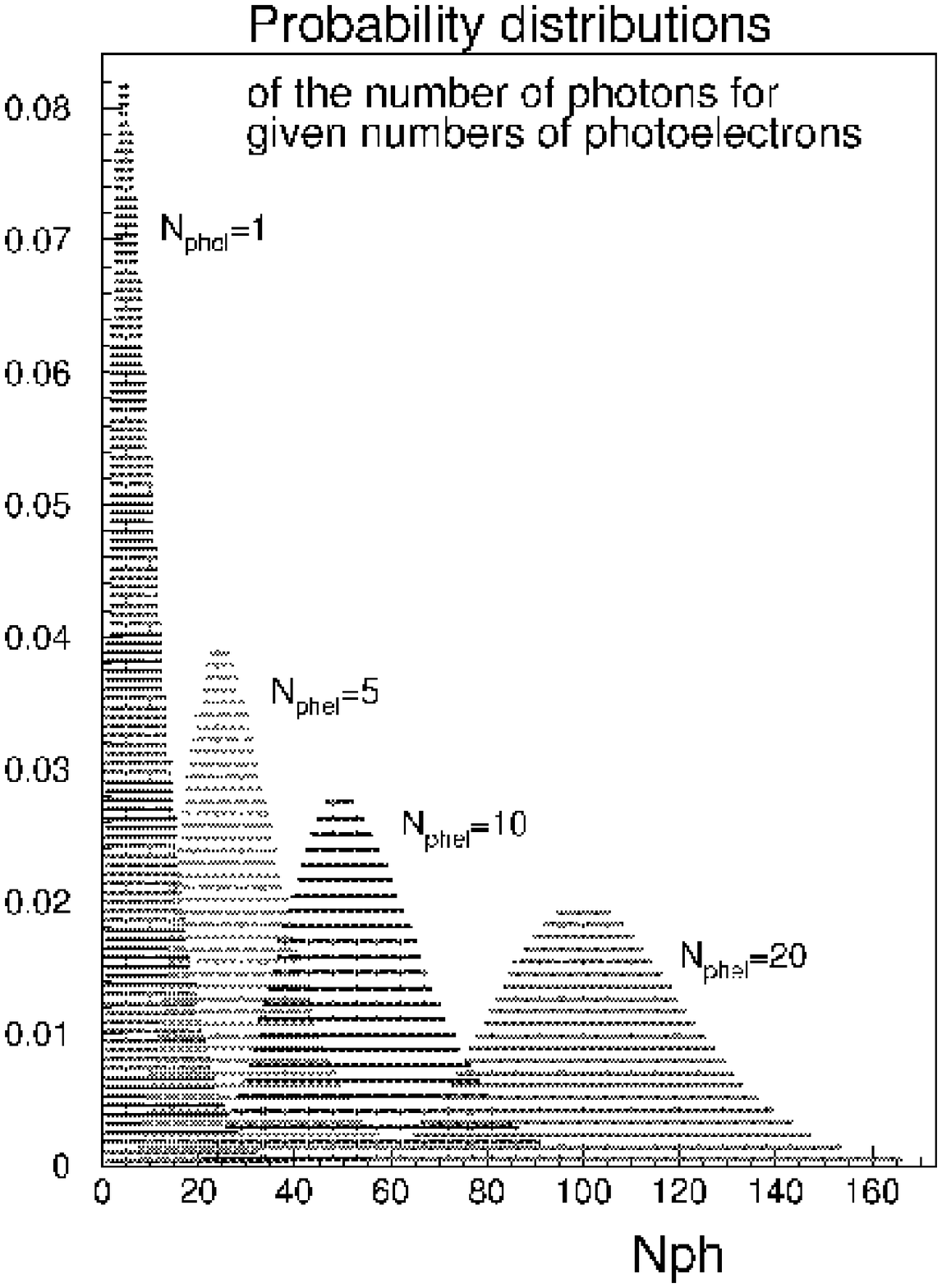}
}
\caption{{fig:6} Probability distributions of the number of the 
Cherenkov photons for given numbers of photoelectron. The number
 of simulation is 100.}

\end{figure}


%

\subsection{Construction of a suitable procedure for the discrimination
of muons from electrons}
\label{sec:3.2}
The examination described in the previous section leads us to conclude
that the probability functions and the approximations used in the
standard SK analysis are not reliable for the discrimination of muons
from electrons.  Therefore, a more reliable procedure, constructed on
a solid theoretical basement, is required.  Here, we will outline a
more reliable and suitable algorithm for discrimination of muons from
the electrons.  
First, we simulate the physical processes for the Cherenkov light as
exactly as possible using GEANT~3.21, and construct new mean
angular structure functions for the Cherenkov light for pattern
recognition of the Cherenkov light due to both the muon and electron.
Second, we derive their relative fluctuation functions for the Cherenkov
light distribution which are critical for constructing an
estimator for particle identification.

  In our mean angular structure functions, we calculate the Cherenkov 
light due to both muon and electrons along the direction of the the 
primary muon and electrons.
Then, followings are considered.\\

(1)  As for the mean longitudinal size of $\mu$-track/electron shower, 
we adopt the total longitudinal length ($T_e$ defined lately) in which 
99.5 percent of the total Cherenkov light are considered.\\
(2) The total longitudinal length along which the particles concerend 
emit the Cherenkov light is emitted is devided into a number of eual 
segments. The number of the division is dependent on the particle type, 
primary energy and the required accuracy of the image presentation.\\
(3) We calculate the mean angular distribution, $F_i^{e,\mu}(\theta)$, 
and its relative fluctuation $\delta_i^{e,\mu}(\theta)$ for each 
segment $i$,and construct the mean angular distribution functions, 
$F^{e,\mu}(\theta,t) \;$ and theri relative fluctuations, 
$\delta^{e,\mu}(\theta,t)$, as function  of radiation angle,$\theta$,
and water layer thickness, $t$. The item(3) will be discussed in a 
subsequent paper\\

We neglect the lateral distributions of particles in both $e$ and
$\mu$ events, i.e., we assume that all Cherenkov light is emitted at
the event axis. This makes our approach invalid for events close and
approximately parallel to the walls of the detector, but it is
adequate for the particle discrimination quality estimates.  The much simpler
approach adopted in the standard SK analysis would result in larger
errors in type and geometry reconstruction procedures, particularly in
case of peripheral events when the longitudinal development of
$e$-shower/$\mu$-track Cherenkov light angular distribution is most
prominent.

Note that angular distribution functions for the Cherenkov light are
universal functions for a given particle type and energy and can be
used for calculating mean pattern images and their deviations for any
required event geometry in any water tank.\\
Further, it should be noticed that we need not introduce 
$P_{pattern}(e(\mu))$ and  
$P_{angle}(e(\mu))$ in the procedure for the discrimination as in Eqs.(7) and
(8) in the final probability function, and directly get 
the final probability function.


\subsection{The construction of the mean angular distribution
 functions of Cherenkov light for muons and electrons}
\label{sec:3.3}
From the view point of pattern recognition, we need more
accurate angular distribution functions for the total Cherenkov light due
to both muons and electrons, rigorously taking into account fluctuations
inherent in both types of event, to confirm the
discrimination between muons and electrons. For this purpose, we
construct angular structure functions for the Cherenkov light in the
following way.

%

\subsubsection{The construction of the mean angular distribution
 function due to a primary electron and its relative fluctuation function.}


An electron due to electron neutrino interaction produces an electron
shower in which shower electrons are distributed over some range in
space, each of which produces Cherenkov light.  It is not
suitable for such a range to be
approximated by the  point-like as assumed in SK, on which we 
have already clarified.
Instead of the SK point-like approximation, we introduce a
`moving-point' approximation in the following to construct the angular
distribution function for the total Cherenkov light.
Here, we construct the mean angular distribution
function for the Cherenkov light for electron.

In a large water tank detector which is large enough for the dimensions
of the electron showers concerned, we calculate the development of total
Cherenkov light due to the electron shower.

We exactly simulate individual angular distributions for the Cherenkov
light using a combination of GEANT~3.21 with the calculation tools 
developed by us. 
To construct the mean angular distribution function for the
Cherenkov light and its relative fluctuation function accurately, we
need to simulate a large number of showers, 10000 to 20000.
Let $T_e$ be the total length of an electron shower
initiated by a primary electron. We divide $T_e$ in segments, the
length of which depends on the primary energy of the charged particle 
concerned and which is taken as 40~cm in present calculation.


  Shower particles are produced according to the exact simulation 
procedure of the electron cascade shower, but in our models of the 
Cherenkov light, both the mean aogular distribution function 
and corresponding angular distribution relative fluctuations 
function are attributed to a certain segment. When calculating 
the pattern images with the help of the models, all Cherenkov 
photons generating within the segment are thought to start 
from the middle of the segment. This approximation is called
the ``moving-point'' (or ``multi-point'') approximation, 
in contrast to the single point approximation adopted in the standard
SK analysis.

We simulate the cascade shower initiated by the primary electron 
and calculate the Cherenkov light from shower particles in 
each segment and obtain the angular
distribution function for the Cherenkov light by distributing all the
Cherenkov photons emitted from segment $k$ over the bins of a
histogram $N_e(\theta,E_0,k)$ in $\theta$ to estimate: \\
{\bf The Mean angular distribution function} \\
\begin{equation}
F_e (\theta_i,E_0,k) \, = \, \frac{\left< N_e(\theta_i,E_0,k)\right >}
{\Delta\Omega_i},
\label{eqn:12}
\end{equation}

where $\left<...\right>$ denotes the  average over a large event sample,
$\theta_i$ is the center of mass of the $i$-th histogram bin, and 
$\Delta\Omega_i$ is the solid angle of the $i$-th bin. \\
{\bf The Relative fluctuation function for the angular distribution} \\
\begin{equation}
\delta_e (\theta_i,E_0,k) \, = \, \frac{\sqrt{\left<N_e^2(\theta_i,E_0,k)
\right> \,- \, \left<N_e(\theta_i,E_0,k)\right>^2}}
{\left<N_e(\theta_i,E_0,k)\right>} \;.
\label{eqn:13}
\end{equation}

Then we fit the mean angular distribution (12) for each segment separately by
a function: \\
\\
\begin{eqnarray}
\lefteqn{F_e (\theta,E_0,k)= } \nonumber \\
&=&
10^{\left \{ \left [ A \, + \, B3/ 
\left ( 1 + B \cdot |\theta - C|^{
\left ( B4 + B1 \cdot |\theta - C| \right )} \right ) \right ] / 
\left [1 \, + \, B2 \cdot \theta
\right ] \right \}}\;, 
\label{eqn:14}
\end{eqnarray}

where the numerical values of A, B, C, B1, B2, B3 and B4 
depend on the primary energy of the electron and the corresponding
segment.

 In Figures~7 to 11, we give the mean angular distribution
function for the total Cherenkov light from each segment for a 500~MeV
electron.  We give the mean angular distribution function for the
total Cherenkov light emitted in segments No.1 (the interval from 0 to
40~cm from the starting point of cascade shower), No.2 (from 40~cm to
80~cm from the starting point), No.3 (80~cm to 120~cm), No.4 (120~cm
to 160~cm), and No.5 (160~cm to 200~cm), 
in Figures~7, 8, 9, 10 and 11, respectively.

\begin{figure}

\resizebox{0.4\textwidth}{!}{%
  \includegraphics{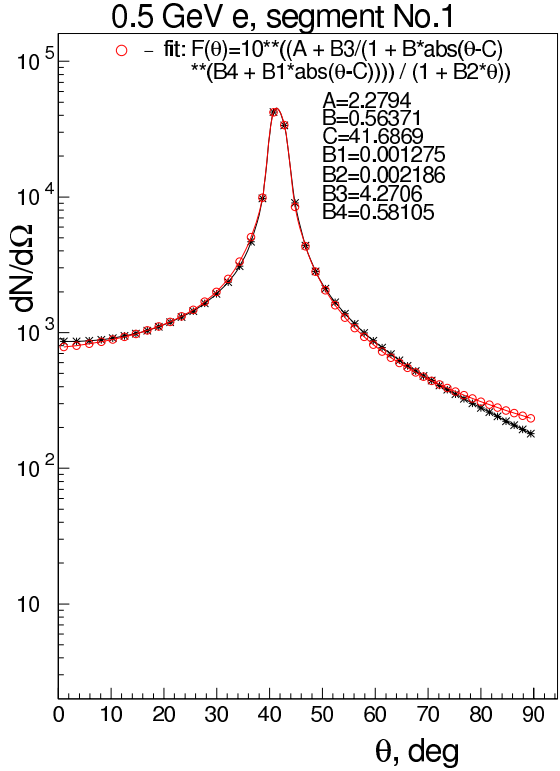}
}
\caption{{fig:7a} The angular structure function for 500 MeV
 electrons from segment No.1. See the text for details.}

\end{figure}

\begin{figure}

\resizebox{0.4\textwidth}{!}{%
  \includegraphics{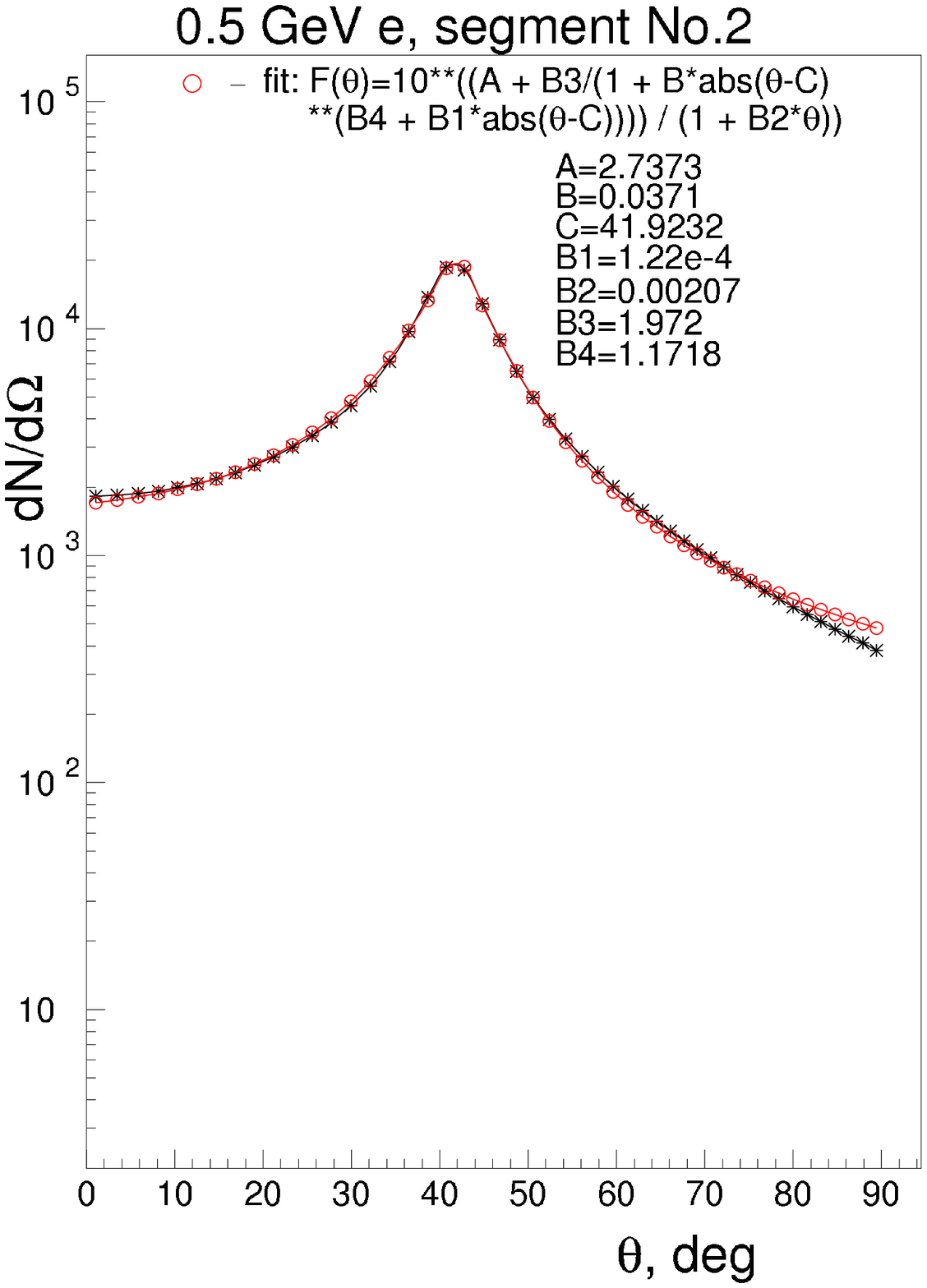}
}
\caption{{fig:7b} The angular structure function for 500 MeV
 electrons from segment No.2. See the text for details.}
\end{figure}

\begin{figure}

\resizebox{0.4\textwidth}{!}{%
  \includegraphics{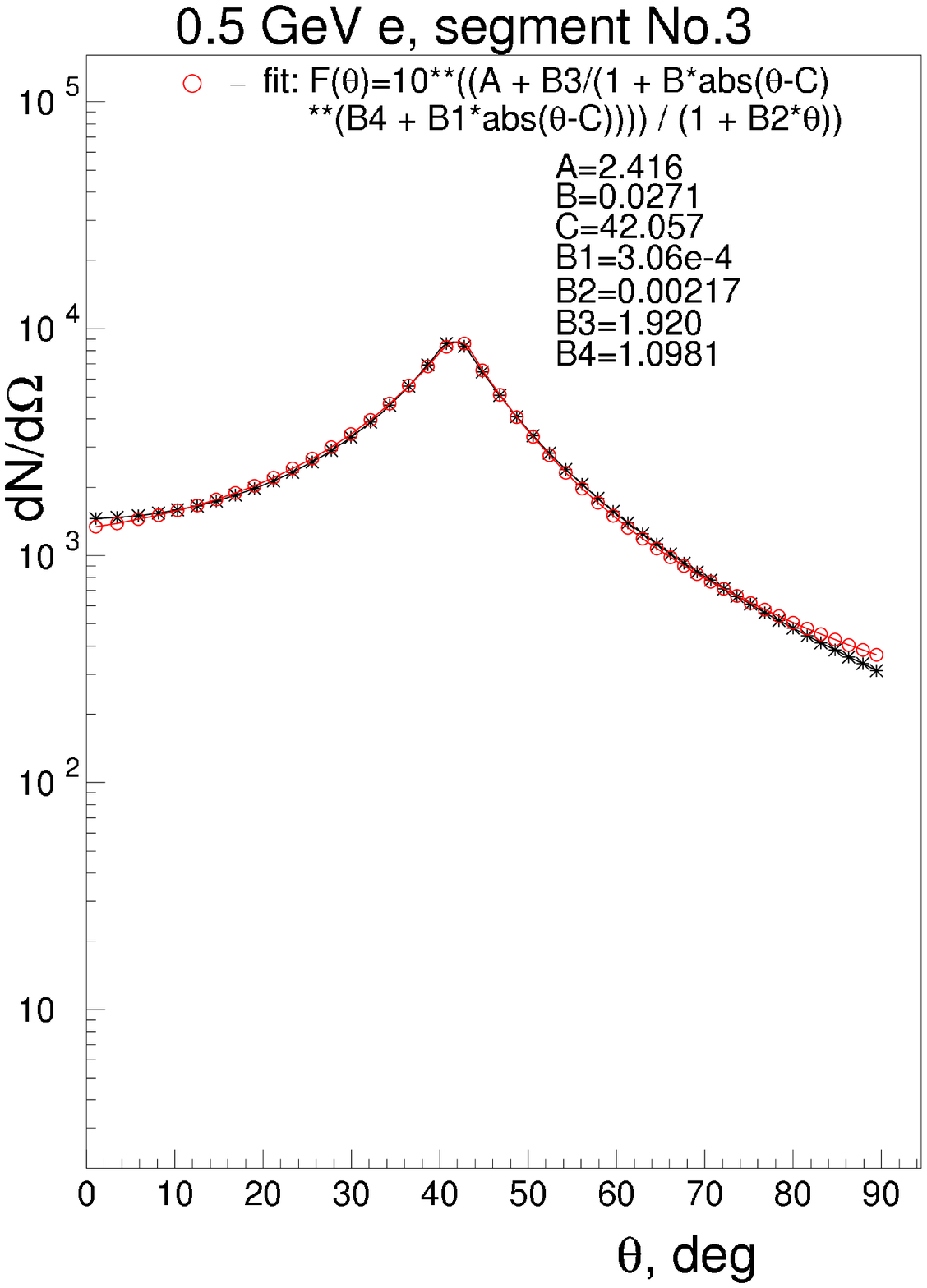}
}
\caption{{fig:7c} The angular structure function for 500 MeV
 electrons from segment No.3. See the text for details.}
\end{figure}

\begin{figure}

\resizebox{0.4\textwidth}{!}{%
  \includegraphics{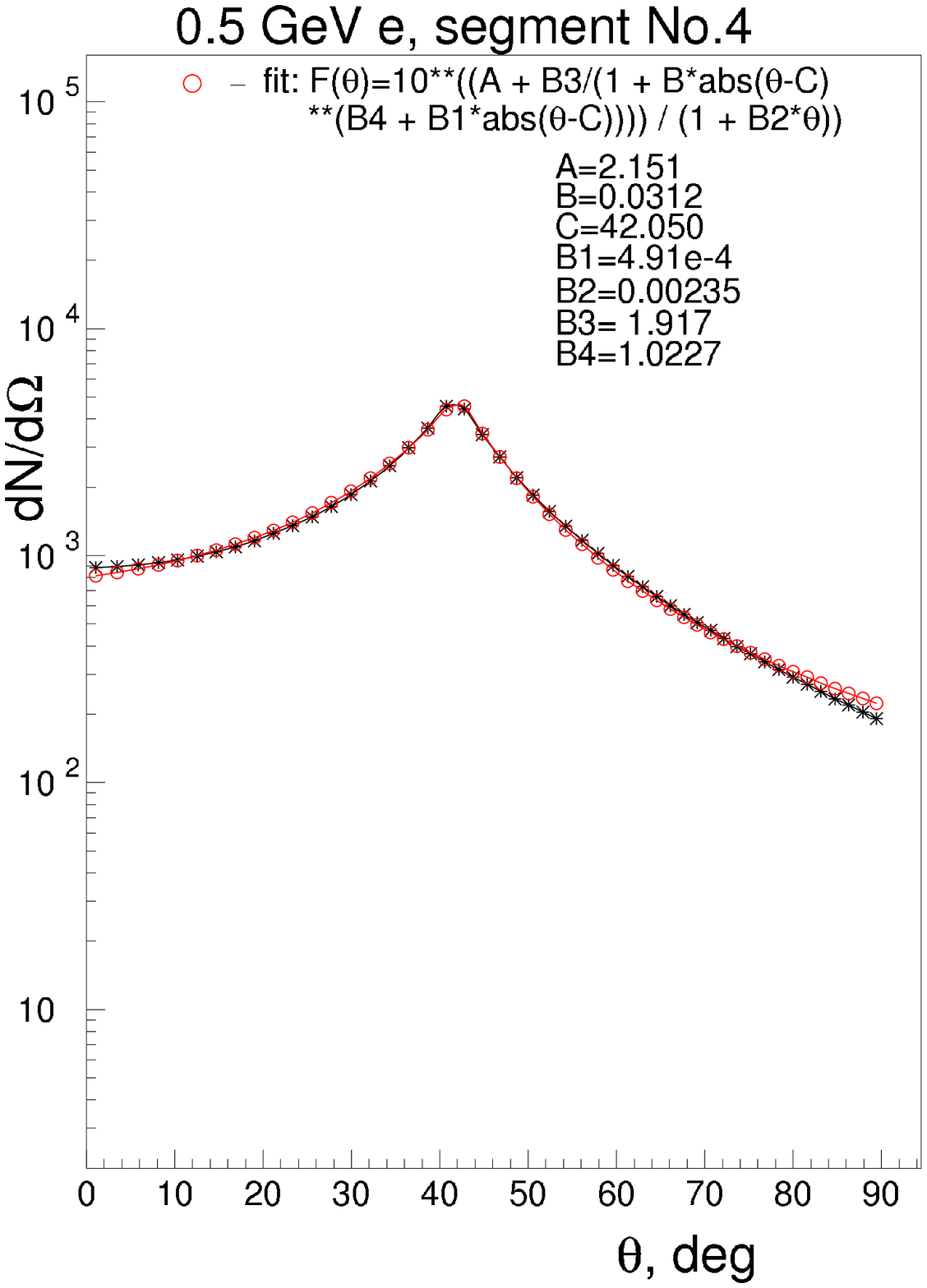}
}
\caption{{fig:7d} The angular structure function for 500 MeV
 electrons from segment No.4. See the text for details.}
\end{figure}

\begin{figure}

\resizebox{0.4\textwidth}{!}{%
  \includegraphics{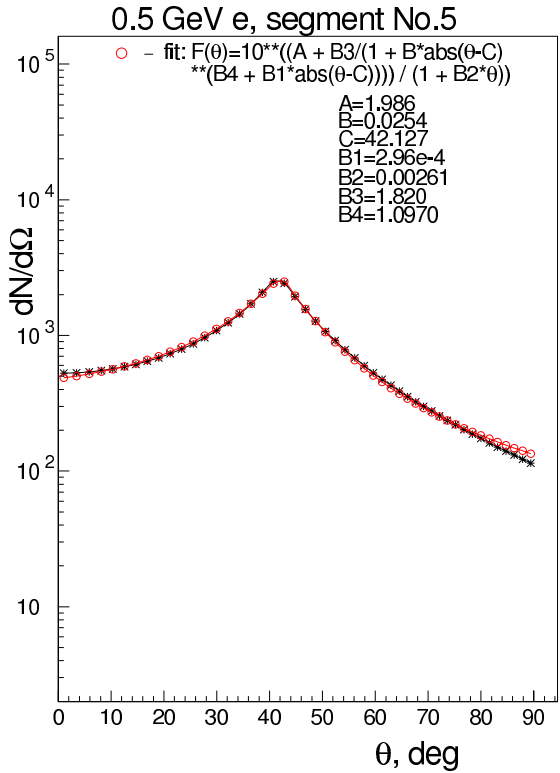}
}
\caption{{fig:7e} The angular structure function for 500 MeV
 electrons from segment No.5. See the text for details.}
\end{figure}

The electron shower produced by a 500~MeV electron is rather small and
the essential character of the cascade shower is determined in
segments~1 to 5.  In Figure~7 (segment
No.1), which corresponds to the initial stage of shower development,
shower particles are energetic and near the core of shower axis and so
give rise to a strong peak around the Cherenkov angle due to energetic
particles initiating the cascade shower.  In Figure~8
(segment No.2), the cascade shower reaches shower maximum, where
number of shower particles reaches a maximum, and the total Cherenkov
light produced in this segment is roughly the same as that in segment
No.1.  However, the average energies of shower particles in this
segment are smaller than those in segment No.1, and they are more
scattered due to multiple scattering and so there is more deviation
from the original Cherenkov angle.

The situation in segment No.3 (Figure~9) is roughly the
same as that in segment No.2(Fig.8).  It should be emphasized that the
amounts of Cherenkov light emitted in segments No.2 and No.3 are
greater than that in the segment No.1 (near the starting point of the
cascade shower) at all angular regions except the original Cherenkov
angle.  Even in the segments No.4 (Figure~10) and No.5(Figure~11),
 where electron shower is attenuated rapidly, the
contributions of Cherenkov light from outside the original Cherenkov
angle could not be neglected compared with that from the segment No.1,
which is too near the starting point of the cascade shower.  In
conclusion, as the Cherenkov light produced in every segment
contributes to the edge of the Cherenkov ring, we cannot say that the
Cherenkov light at the edge of the Cherenkov ring comes exclusively
from the starting point of the cascade shower (the vertex point of the
electron neutrino reaction), which is the assumption adopted in the SK
analysis (See Figure~2.9 in Sakai,\cite{Sakai}).  
Here, we have paid particular attention to
a 500~MeV electron, however, such characteristics hold irrespective of
the primary energy of the electron.

In order to calculate the relative fluctuation function for the
angular distribution accurately, a large sample of simulated events is
needed: we used 20000 events for $E_0 \le$500~MeV and 10000 events for
$E_0 >$500~MeV.

 In Figures~12 to 16 we construct the relative fluctuation function for the
Cherenkov light from a 500~MeV electron. The relative fluctuation
functions are extremely important for the pattern recognition of
electrons and muons. Therefore, we should determine them as accurate
as possible, which was why we simulated such a large number of events.

In Figures~12, 13, 14, 15 and 16 we give the relative
fluctuations for the Cherenkov light from segments 1, 2, 3, 4 and 5,
respectively.

\begin{figure}
\resizebox{0.4\textwidth}{!}{%
  \includegraphics{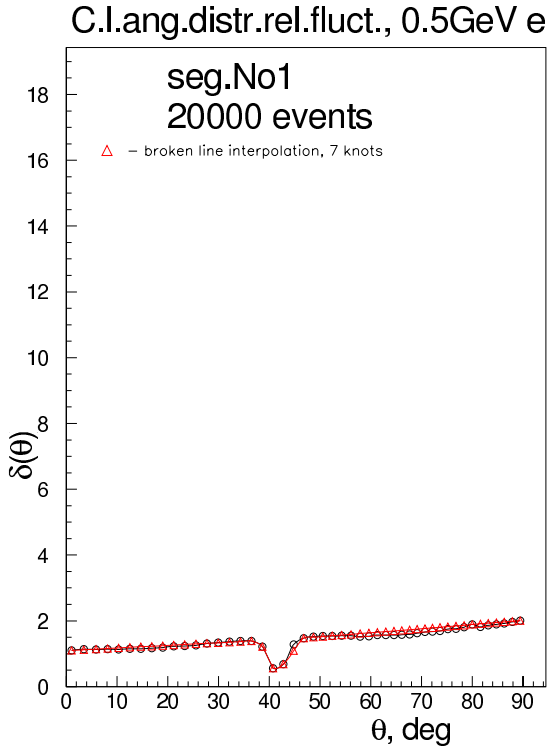}
}
\caption{{fig:8a} The relative fluctuations in the angular
 distribution of the Cherenkov light for 500 MeV electrons 
from segment No.1. See the text for details.}
\end{figure}

\begin{figure}
\resizebox{0.4\textwidth}{!}{%
  \includegraphics{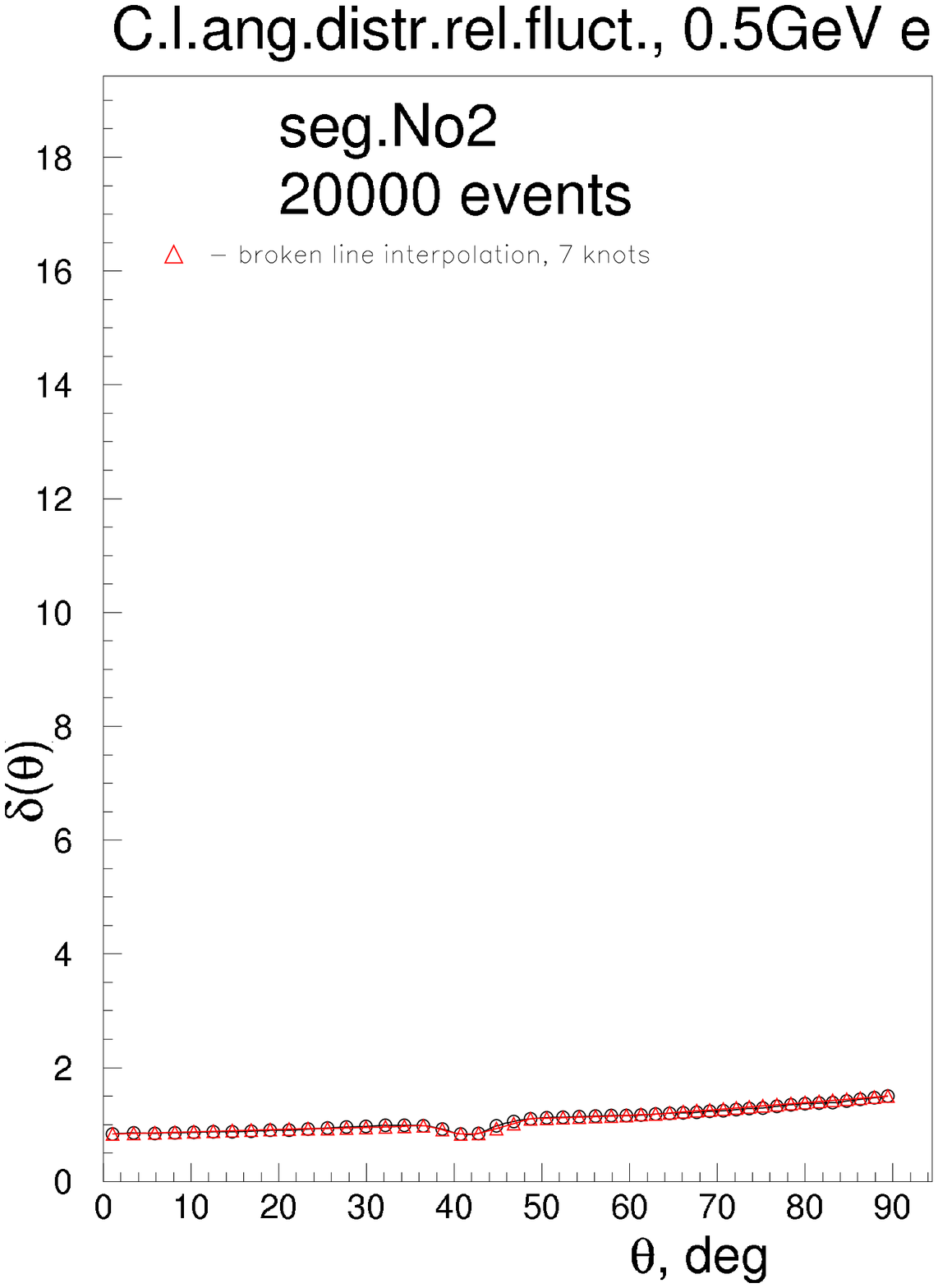}
}
\caption{{fig:8b} The relative fluctuations in the angular
 distribution of the Cherenkov light for 500 MeV electrons 
from segment No.2. See the text for details.}
\end{figure}  

\begin{figure}
\resizebox{0.4\textwidth}{!}{%
  \includegraphics{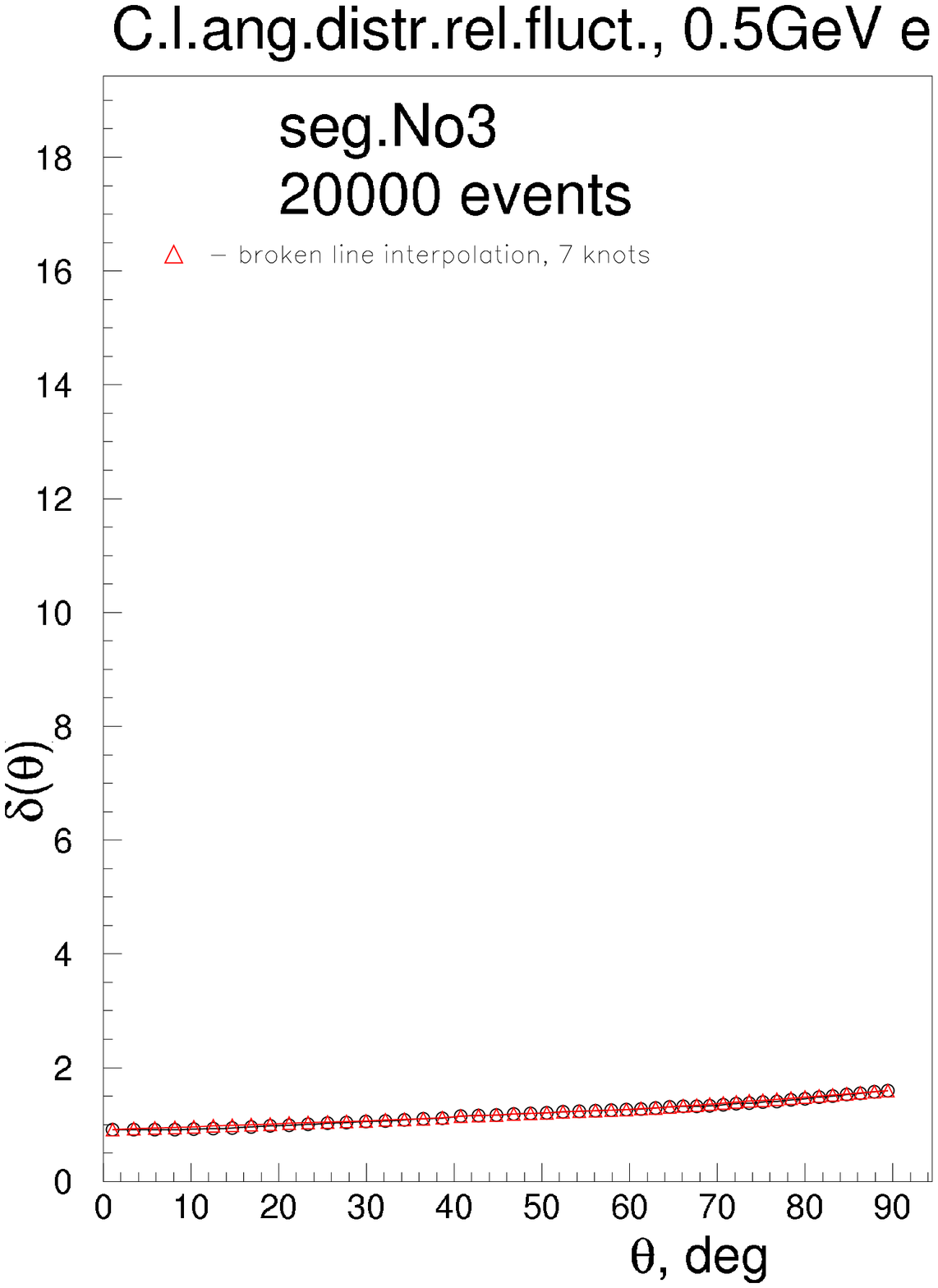}
}
\caption{{fig:8c} The relative fluctuations in the angular
 distribution of the Cherenkov light for 500 MeV electrons 
from segment No.3. See the text for details.}
\end{figure}

\begin{figure}
\resizebox{0.4\textwidth}{!}{%
  \includegraphics{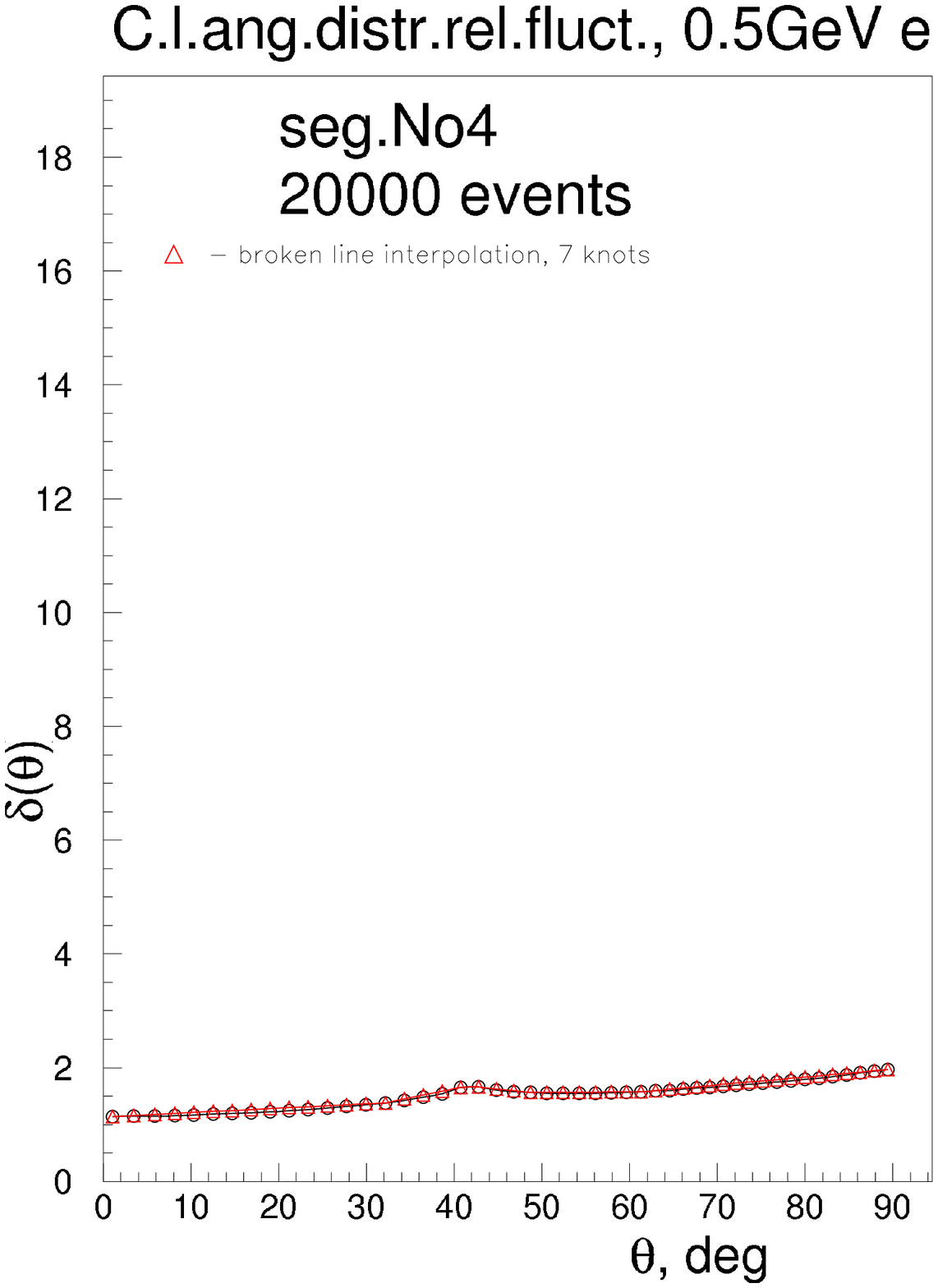}
}
\caption{{fig:8d} The relative fluctuations in the angular
 distribution of the Cherenkov light for 500 MeV electrons 
from segment No.4. See the text for details.}
\end{figure}

\begin{figure}
\resizebox{0.4\textwidth}{!}{%
  \includegraphics{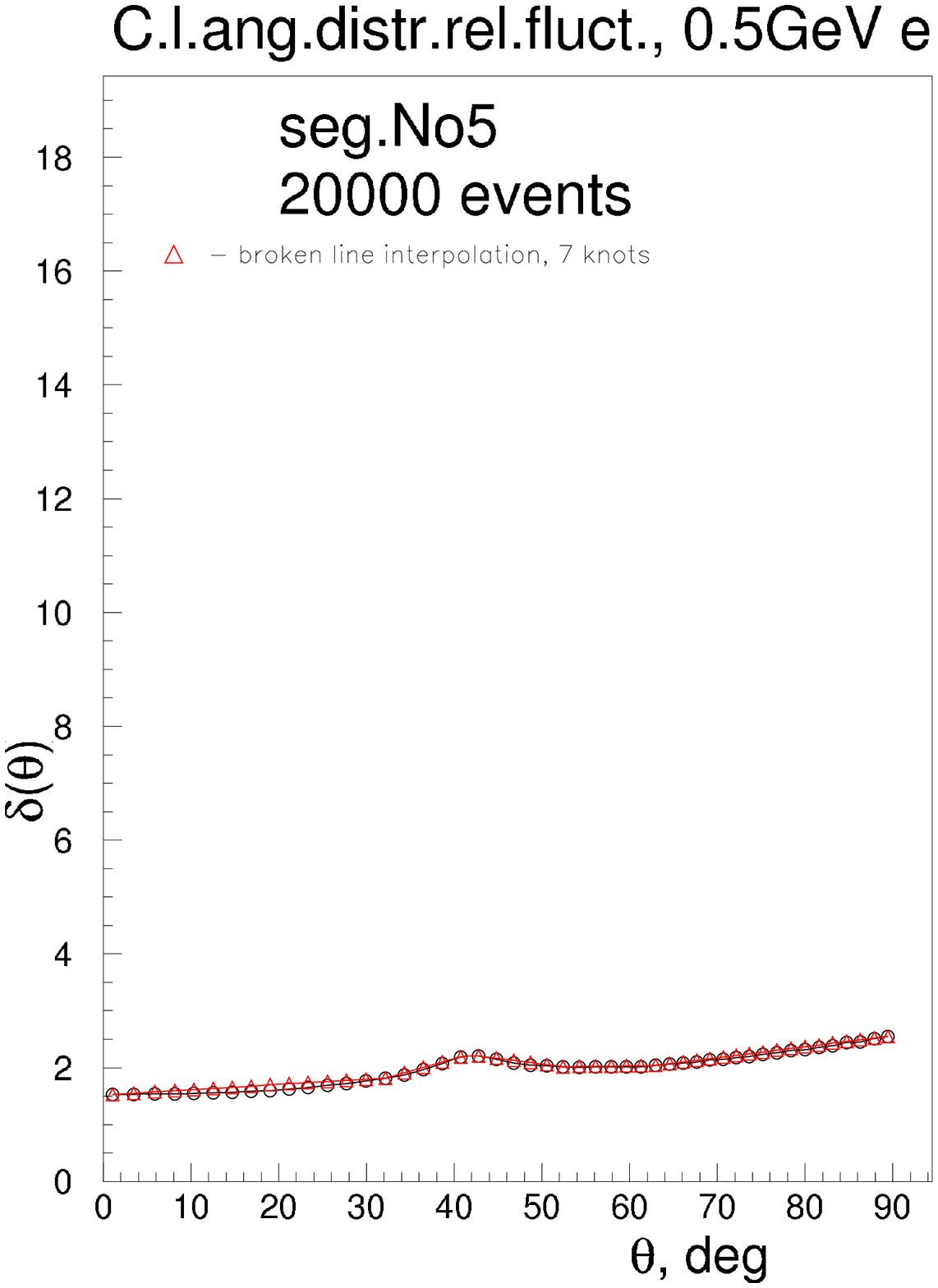}
}
\caption{{fig:8e} The relative fluctuations in the angular
 distribution of the Cherenkov light for 500 MeV electrons 
from segment No.5. See the text for details.}
\end{figure}

In Figure~12 (segment No.1), the shower particles are
relatively energetic so that the relative fluctuation is a minimum
near 42$^\circ$, the original Cherenkov angle, as expected.  In
Figure~13 (segment No.2), the situation is the same as in
Figure~12.  It is interesting to note that the minimum
disappears in the segment No.3(Fig. 14) and a maximum appears at the original
Cherenkov angle in segments No.4(Fig.15) and No.5(Fig.16).  However, globally
speaking, the angular dependence of the fluctuation is weak and it is
relatively similar at all angles.  Of course, this does not mean that
shower particles are produced isotropically, losing the direction of
the initiating particle.

The fluctuations in the electron events which are described above are
quite different from those in muon events (see next subsection).
  
We need to take these
relative fluctuations into account for more reliable discrimination
between electrons and muons. This is the reason why we calculate the
relative fluctuations after simulating each energy 10000 to 20000
times.




\subsubsection{The construction of the mean angular distribution
 function due to a primary muon and its relative fluctuation function.}



In the same way as in the preceding sub-section, we obtain
the mean angular distribution function $F_{\mu}(\theta,E_0,k)$ for the
Cherenkov light emitted due to the primary muon and its relative
fluctuation $\delta_{\mu}(\theta,E_0,k)$.

We then fit the mean angular distribution for each segment in turn by a
function: \\ 
\begin{equation}
F_{\mu}(\theta,E_0,k) \, = \, 10^{\left \{ A \, exp \left [ -B \, 
(\theta - C)^2 \right ]
\right \}} \, + \,  10^{\left [ B1/(1+B2 \, \theta^4) \right ]}
 \;, 
\label{eqn:15} 
\end{equation}
where the numerical values of A, B, C, B1 and B2 are dependent
on both $E_0$, the primary energy of muon, and $k$, the segment.

\begin{figure}
\resizebox{0.4\textwidth}{!}{%
  \includegraphics{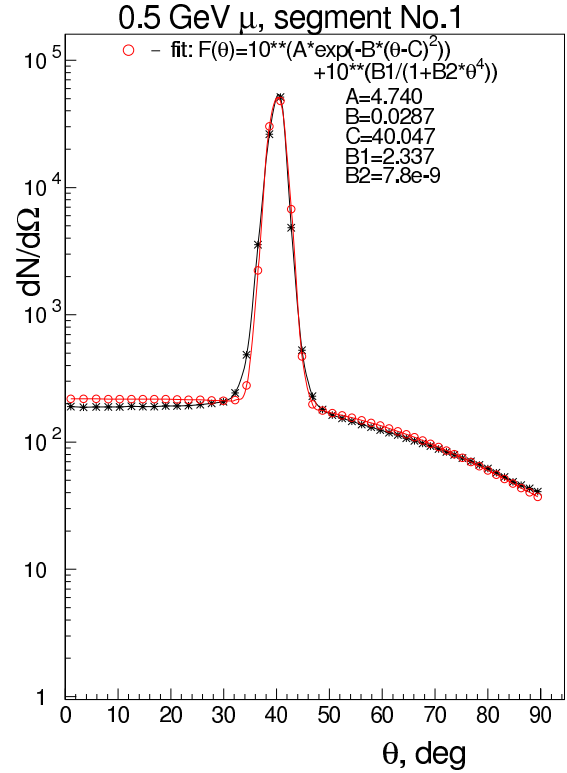}
}
\caption{{fig:9a}  The angular structure function for 500 MeV 
muons from segment No.1. See the text for details.}
\end{figure}

\begin{figure}
\resizebox{0.4\textwidth}{!}{%
  \includegraphics{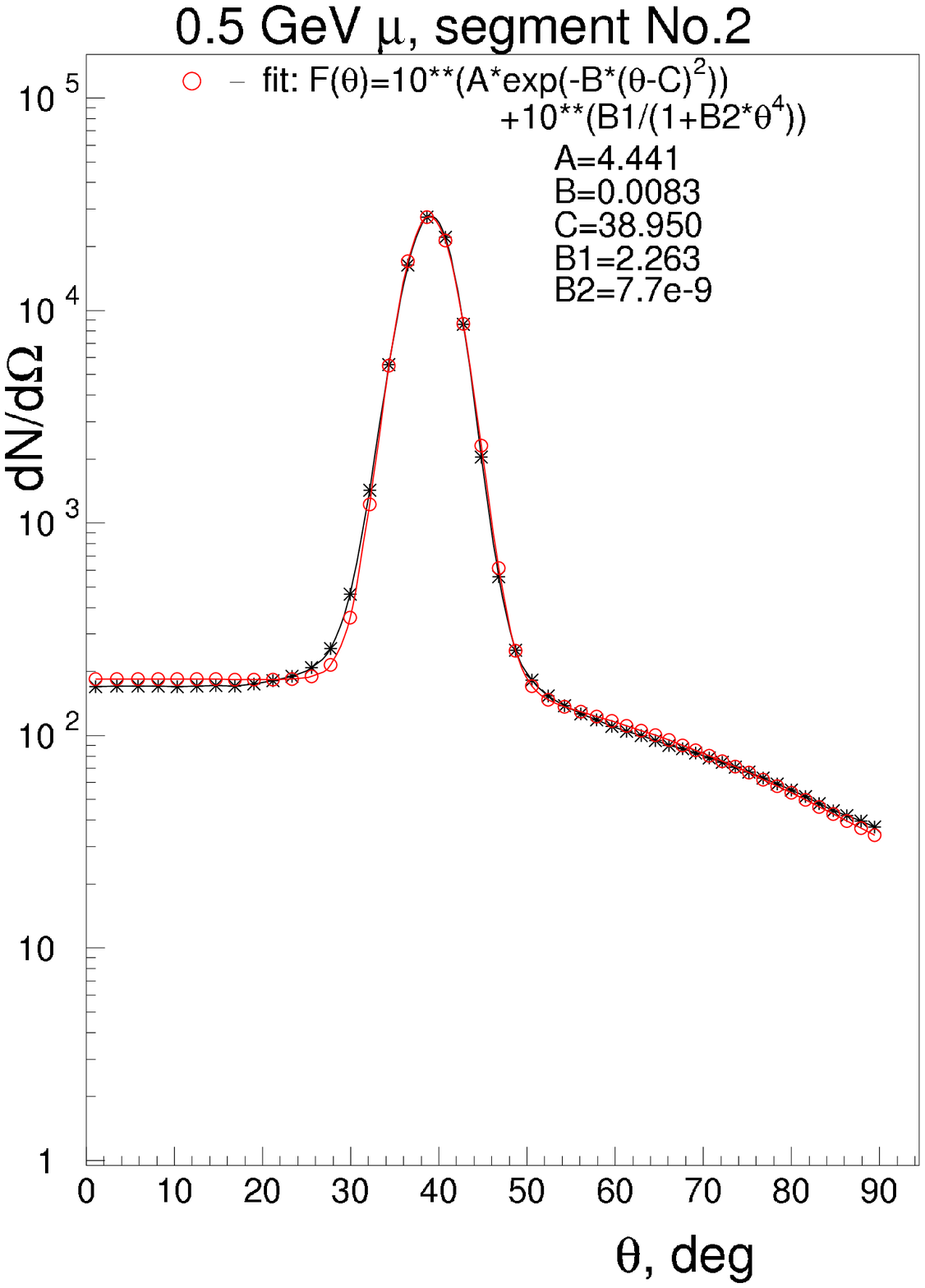}
}
\caption{{fig:9b}  The angular structure function for 500 MeV 
muons from segment No.2. See the text for details.}
\end{figure}

\begin{figure}
\resizebox{0.4\textwidth}{!}{%
  \includegraphics{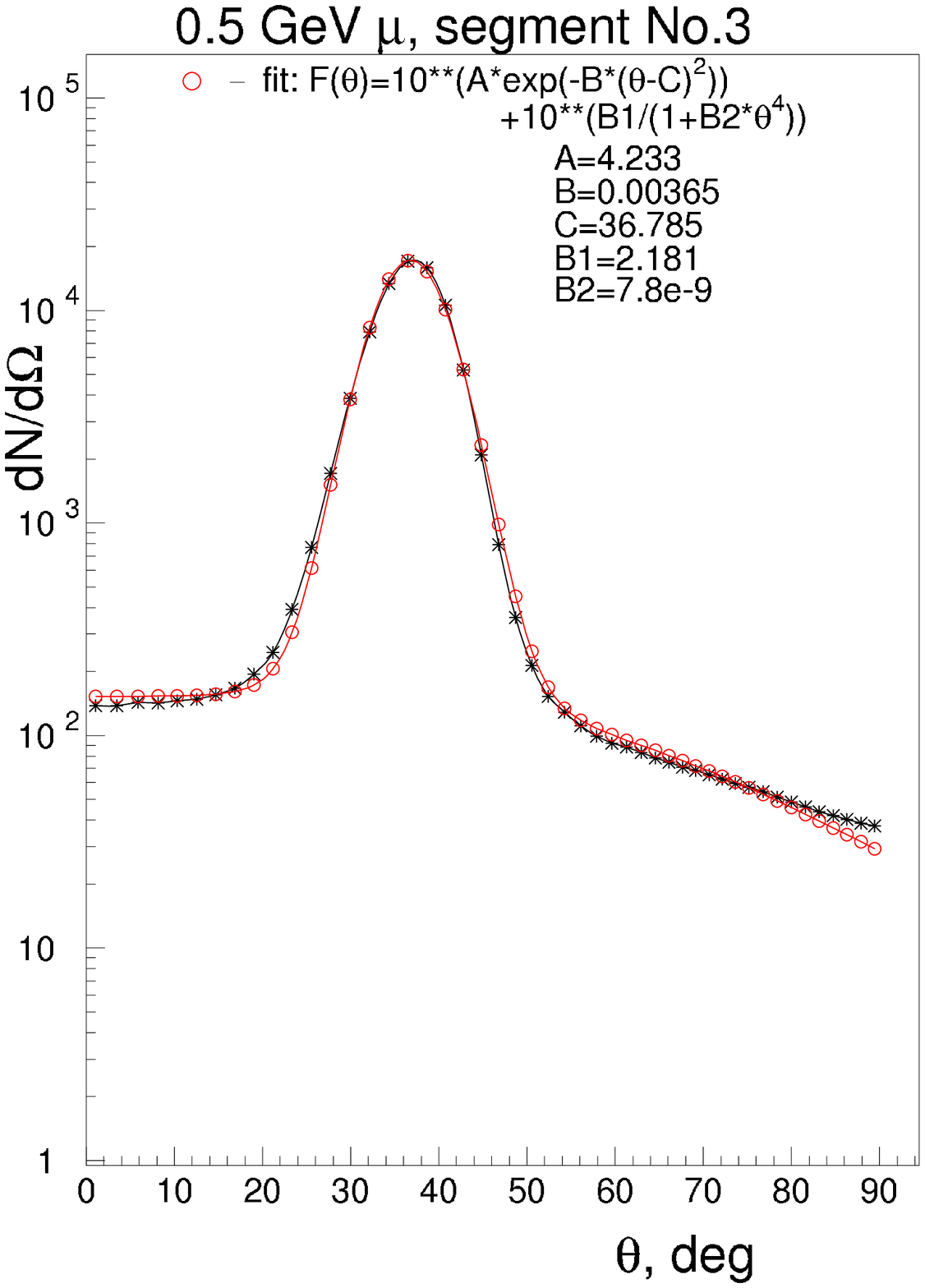}
}
\caption{{fig:9c}  The angular structure function for 500 MeV 
muons from segment No.3. See the text for details.}
\end{figure}

\begin{figure}
\resizebox{0.4\textwidth}{!}{%
  \includegraphics{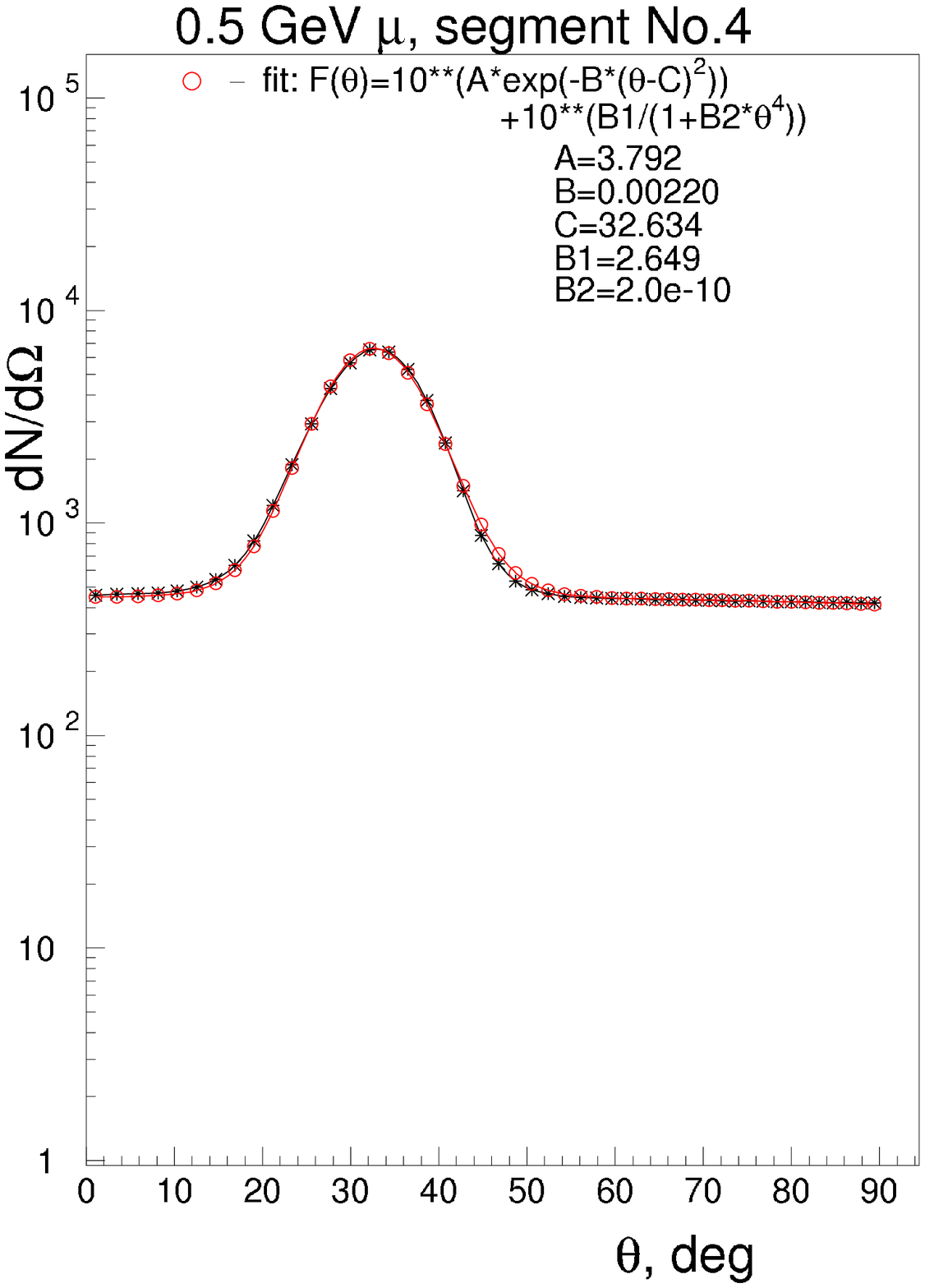}
}
\caption{{fig:9d}  The angular structure function for 500 MeV 
muons from segment No.4. See the text for details.}
\end{figure}

\begin{figure}
\resizebox{0.4\textwidth}{!}{%
  \includegraphics{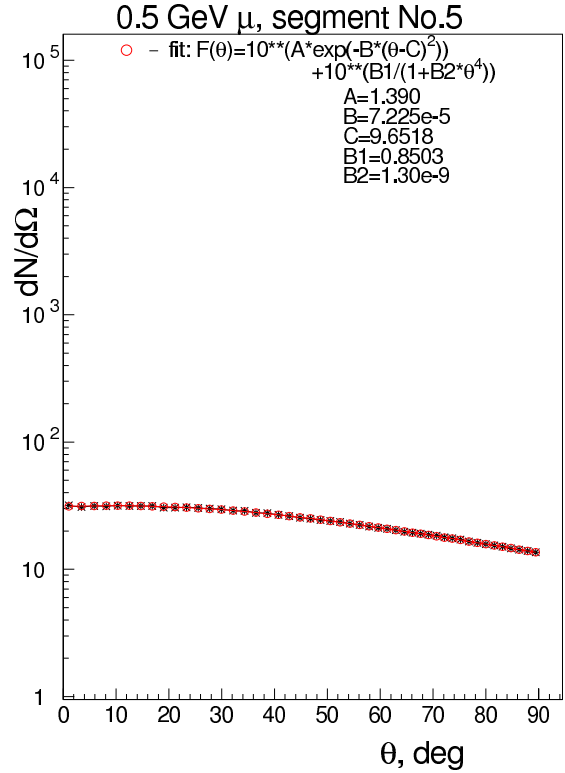}
}
\caption{{fig:9e}  The angular structure function for 500 MeV 
muons from segment No.5. See the text for details.}
\end{figure}

In Figures~17 to 21 we construct the mean angular distribution function for
the total Cherenkov light from each segment for a 500~MeV muon.  We
give the mean angular distribution function for the total Cherenkov
light due to segments No.1 (the interval from 0 to 40~cm from the
starting point of cascade shower)(Fig.17), No.2 (40~cm to 80~cm)
(Fig.18), No.3 (80~cm to 120cm)(Fig.19), No.4 (120~cm to 160~cm)
(Fig.20) and No.5 (160~cm to 200~cm) (Fig.21)  Comparing
Figure~17 with Figure~7, one can easily understand that
 the peak
around 42$^\circ$ in the case of the muon is very sharp compared with
that in the case of the electron.

This means that the larger contribution of the Cherenkov light is due
to muon itself and the contributions to either side of the peak are
from knock-on electrons and others with lower energies in the case of
a muon.  In the case of the electron initiated shower, the cascade of
electrons with different energies and directions produces a broader
spread of Cherenkov light.  As the muon proceeds it loses energy, so
that the angular distribution for the Cherenkov light becomes wider by
multiple scattering while its mean value decreases due to the
reduction of the Cherenkov angle.  We can see this in Figures~18,
19 and 20, where the angular distributions change greatly between
segments, while the production of Cherenkov light is maintained, which
is a different characteristic from the corresponding curves in the
case of the electron.  In Figure~21, we find the disappearance of
the peak of the angular distribution which was evident in the
preceding segments.  The uniformity of the distribution in Figure~21
shows that the decay electrons are distributed isotropically in this
segment due to the nature of the three-body decay.  According to the
 calculations on transition curves for the differential and integral 
 Cherenkon photons, it is easily understood that there is no muon in 
 segment No.5(Fig.21)
(160~cm to 200~cm from the starting point of the muon).

Also, we could not infer the starting point of the muon (the vertex
point of muon neutrino interaction) from the edge of the Cherenkov
ring, as adopted by the SK group, just as for the electron event.
Comparing Figure~17 with Figures~18,19,20 the contributions from 
the segments No.2 (Fig.18), No.3(Fig.19) and No.4(Fig.20) near the edge of
the Cherenkov ring are comparable with that from the segment No.1.  In
conclusion, we can determine the vertex point of the neutrino
interaction, or the direction of the incident neutrino, based on the
total pattern of Cherenkov light only, and not solely from the edge of
the Cherenkov ring.

\begin{figure}
\resizebox{0.4\textwidth}{!}{%
  \includegraphics{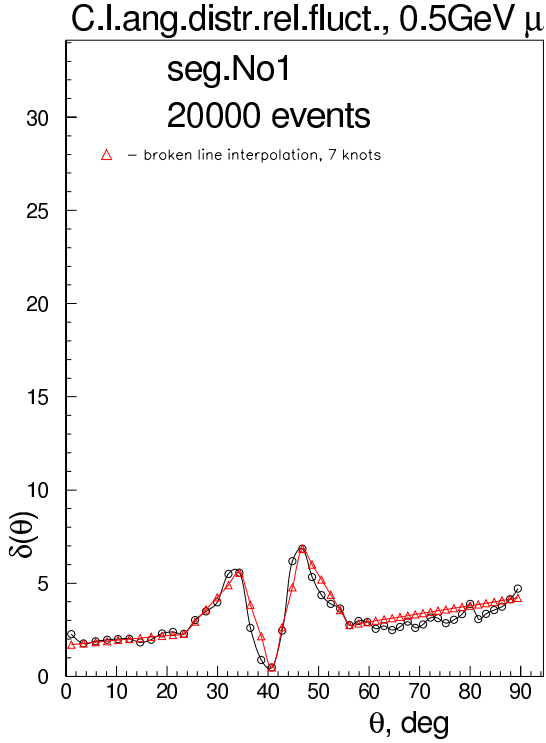}
}
\caption{{fig:10a}  The relative fluctuations in the angular
 distribution of the Cherenkov light for 500 MeV muons from
 segment No.1. See the text for details.}
\end{figure}

\begin{figure}
\resizebox{0.4\textwidth}{!}{%
  \includegraphics{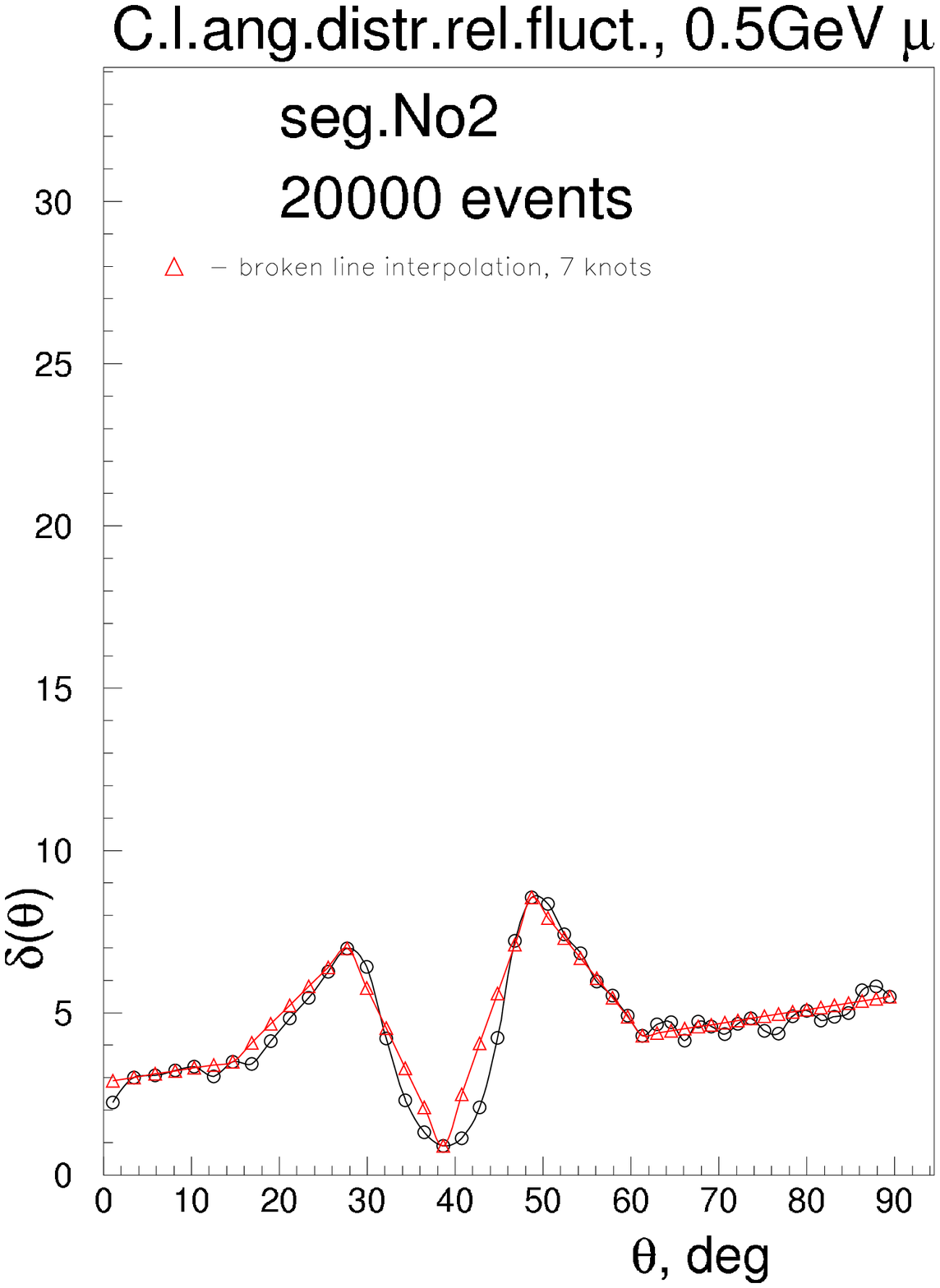}
}
\caption{{fig:10b}  The relative fluctuations in the angular
 distribution of the Cherenkov light for 500 MeV muons from
 segment No.2. See the text for details.}
\end{figure}

\begin{figure}
\resizebox{0.4\textwidth}{!}{%
  \includegraphics{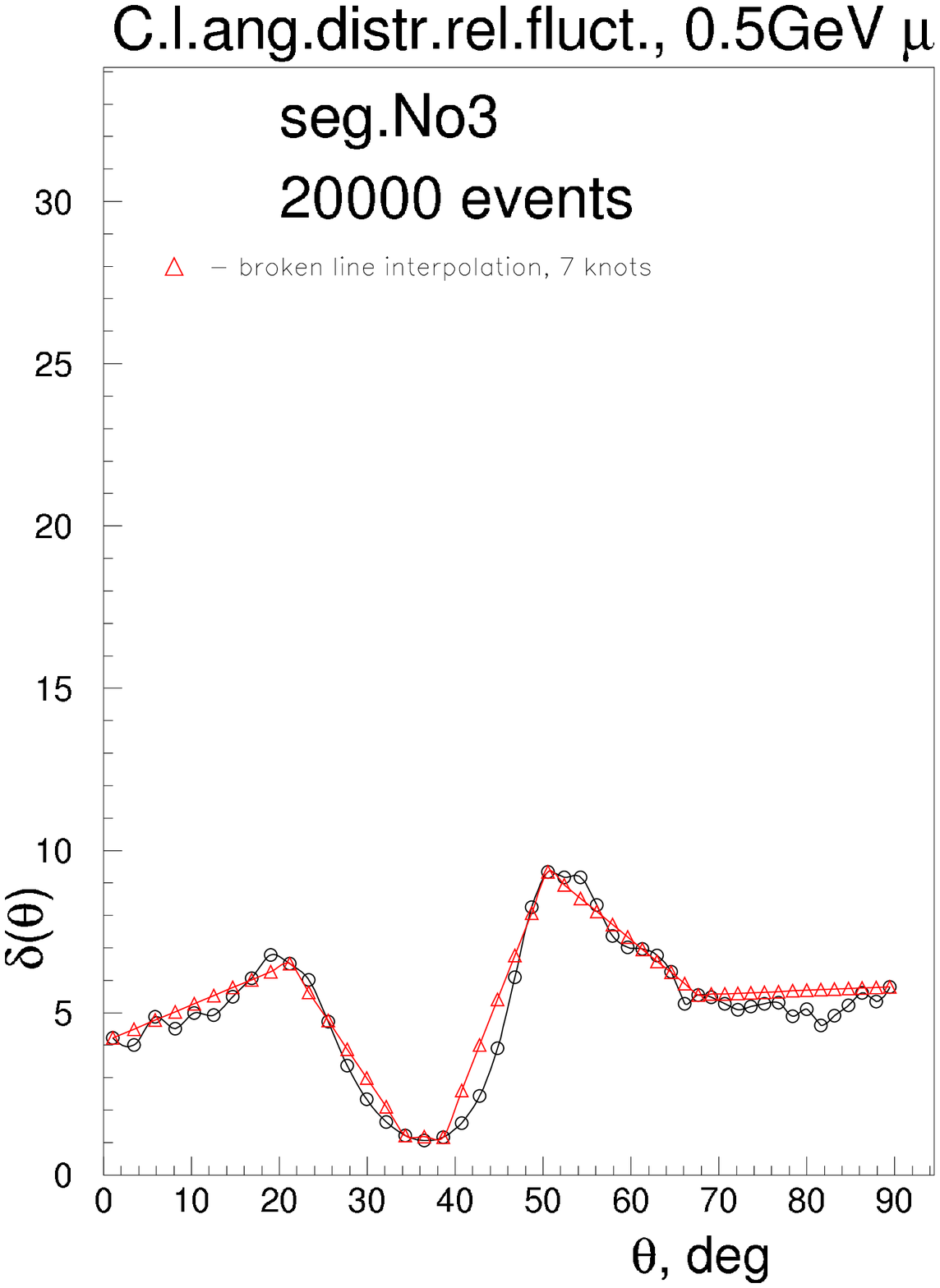}
}
\caption{{fig:10c}  The relative fluctuations in the angular
 distribution of the Cherenkov light for 500 MeV muons from
 segment No.3. See the text for details.}
\end{figure}

\begin{figure}
\resizebox{0.4\textwidth}{!}{%
  \includegraphics{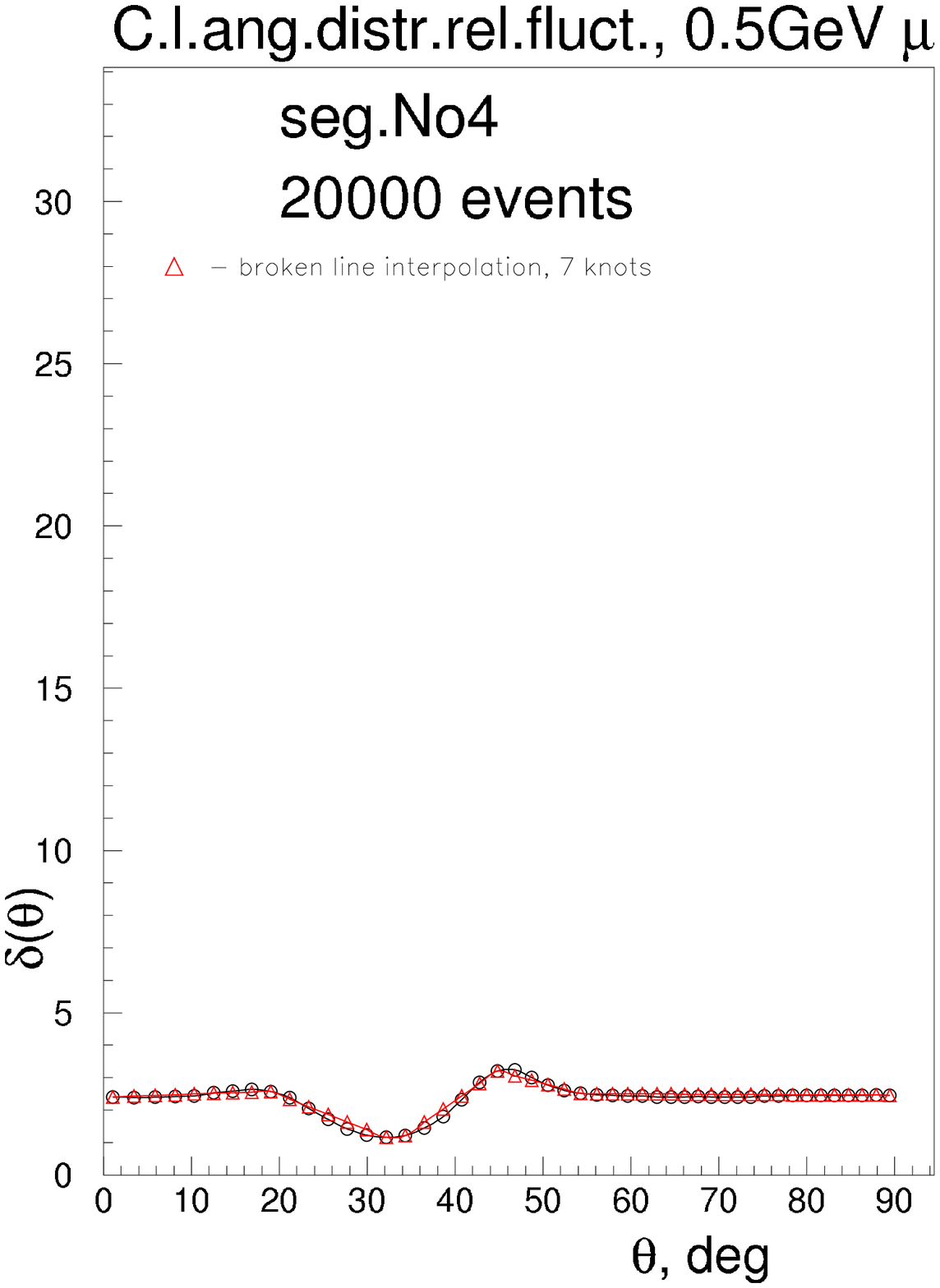}
}
\caption{{fig:10d}  The relative fluctuations in the angular
 distribution of the Cherenkov light for 500 MeV muons from
 segment No.4. See the text for details.}
\end{figure}

\begin{figure}
\resizebox{0.4\textwidth}{!}{%
  \includegraphics{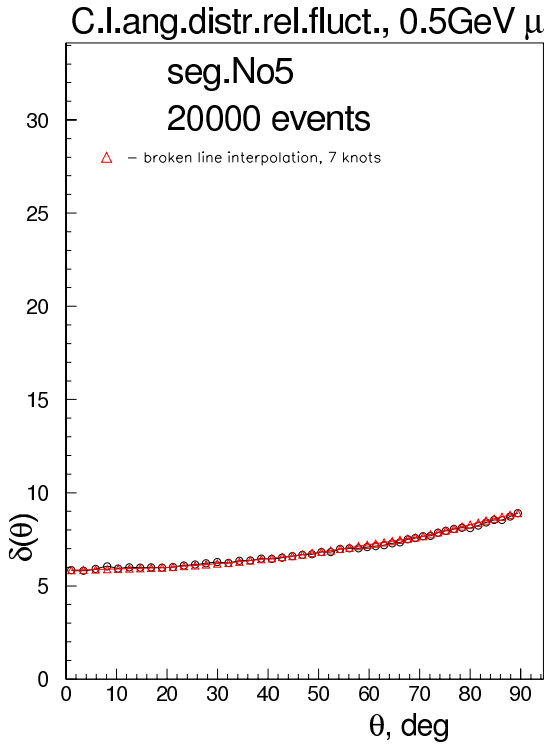}
}
\caption{{fig:10e}  The relative fluctuations in the angular
 distribution of the Cherenkov light for 500 MeV muons from
 segment No.5. See the text for details.}
\end{figure}

We can see that there is a clear difference in the relative
fluctuations between the muon and electrons as well as that in the
angular distribution functions. In Figures~22 to 25, there are
peaks in the relative fluctuation either side of the local minimum at
42$^\circ$. More exactly speaking, the strong concentration of the
Cherenkov light around 42$^\circ$, which is the consequence of the
large contribution by the muon gives a deep minimum to
fluctuation. The two peaks appear as the result of this, and the local
minimum shifts to smaller angles, accompanied by the two peaks, as the
muon proceeds, while there is only one peak in the case of an
electron-initiated cascade.

However, in Figure~26 where the muon decays completely
in the segment No.5 (160~cm to 200~cm from the starting point of the muon),
the (decay product) electron is, on average,
distributed uniformly and the distribution of the Cherenkov becomes
uniform, so that the relative fluctuation also becomes uniform.

Comparing Figure~22 to 26 with Figure~12 to 16, it is easily understood 
that we could expect larger fluctuation in muon events than in electron
events.  The reasons are as follows: In electron events, the shower
particles produced compensate for the effects of fluctuations, while
in the muon events the single muon concerned bears the effect of the
fluctuation exclusively due to multiple scattering, which are not
smeared out.  As a result, we could expect a larger uncertainty in
muon events for both the vertex position and the direction of the
neutrino than in electron events.  We discuss this problem in a 
subsequent paper.


\section{Acknowledgement}


One of the authors (V.G.) should like to thank Prof.\ M.\ Higuchi,
Tohoku Gakuin University. Without his invitation, V.G.\ could not join
in this work.  Authors would like to be very grateful for the
remarkable improvement of the manuscript to Dr. Philip Edwards.

\end{document}